\documentclass[12pt]{article}

\usepackage{amsmath}
\usepackage{amssymb}
\usepackage[utf8]{inputenc}

\numberwithin{equation}{section}
\numberwithin{table}{section}

\newcommand{\dal}{{\dot\alpha}}
\newcommand{\dalpha}{{\dot\alpha}}
\newcommand{\bth}{\bar{\theta}}

\newcommand{\thz}{\theta\theta}
\newcommand{\thq}{\bar{\theta}\bar{\theta}}
\newcommand{\thvoll}{\theta\theta\bar{\theta}\bar{\theta}}
\newcommand{\thsigmath}{\theta\sigma^{m}\bar{\theta}}
\newcommand{\idtdtq}{\int\!\! d^{2}\!\theta d^{2}\!\bar{\theta}\,}
\newcommand{\ifull}{\int\!\! d^4\!x \! \int\!\! d^{2}\!\theta d^{2}\!\bar{\theta}\,}
\newcommand{\idt}{\int\!\! d^{2}\!\theta\,}
\newcommand{\idtq}{\int\!\! d^{2}\!\bar{\theta}\,}

\newcommand{\ivoll}{\int\!\! d^{8}\!z\,}
\newcommand{\ix}{\int\!\! d^{4}\!x\,}
\newcommand{\Lag}{{\mathcal{L}}}
\renewcommand{\Im}{\text{Im}}
\renewcommand{\Re}{\text{Re}}
\newcommand{\hc}{\text{h.c.}}

\newcommand{\bal}{\begin{aligned}}
\newcommand{\eal}{\end{aligned}}
\newcommand{\der}[2]{\frac{\del #1}{\del #2}}
\newcommand{\dder}[3]{\frac{\del^2 #1}{\del #2 \del #3}}

\DeclareMathOperator{\Hess}{Hess}
\def\tn#1{\textnormal{#1}}

\def\del{\partial}
\def\veps{\varepsilon}
\def\=>{\Rightarrow}
\def\<=>{\Leftrightarrow}
\def\l{\left}
\def\r{\right}

\def\btheta{\bar\theta}

\def\hK{\hat{K}}
\def\hG{\hat{G}}
\def\ba{{\bar a}}
\def\bb{{\bar b}}
\def\bc{{\bar c}}
\def\bd{{\bar d}}
\def\bi{{\bar{i}}}
\def\bj{{\bar{j}}}
\def\cc{\hat{c}}

\usepackage{cite}

\begin{document}
\begin{titlepage}
\begin{center}
\rightline{\small ZMP-HH/12-10}
\vskip 1cm

{\Large \bf
Duality and Couplings of 3-Form-Multiplets\\[1.5ex]
in $N=1$ Supersymmetry}
\vskip 1.2cm
{\bf Kai Groh$^a$, Jan Louis$^{b,c}$, Jason Sommerfeld$^b$}

\vskip 0.8cm

$^a${\em School of Mathematical Sciences, University of Nottingham, Nottingham, UK}
\vskip 0.4cm
$^{b}${\em II. Institut f\"ur Theoretische Physik der Universit\"at Hamburg, Luruper Chaussee 149, 22761 Hamburg, Germany}
\vskip 0.4cm
{}$^{c}${\em Zentrum f\"ur Mathematische Physik,
Universit\"at Hamburg,\\
Bundesstrasse 55, D-20146 Hamburg, Germany}
\vskip 0.8cm

{\tt jan.louis@desy.de, kai.groh@nottingham.ac.uk, jason.sommerfeld@desy.de}

\end{center}

\vskip 1cm

\begin{center} {\bf ABSTRACT } \end{center}

\noindent
In this paper we study 3-form gauge fields
in four-dimensional ${N=1}$ supersymmetric theories.  We give the sigma model action together with its Poincar\'e dual action for massless and massive 3-forms. 
The resulting target space geometries are  K\"ahler where the respective 
field variables and superspace couplings are
related  by a Legendre transformation.

\bigskip

\vfill

December 2012

\end{titlepage}

\tableofcontents
\nocite{*}
\newpage

\section{Introduction}
Supersymmetric multiplets which contain $p$-form gauge fields are of
particular interest in string theory  as they
generically appear in the low energy effective action.
In standard Calabi-Yau compactifications these $p$-forms are massless
while in  generalized compactifications 
and/or string backgrounds with non-trivial background fluxes 
they can be massive gaining their mass via a St\"uckelberg 
mechanism \cite{Polchinski:1995sm,Louis:2002ny,Grimm:2004uq}.

In this paper we study massless and massive 3-form gauge fields $C_{npq}$
in  four-dimen\-sional ($D=4$) $N=1$ globally supersymmetric theories.
When $C_{npq}$ is massless its equation of motion forces the field strength 
to be a constant and thus no propagating
degree of freedom is left \cite{fieldreps}.
For this reason massless 3-forms have also been discussed in connection with
the problem of the cosmological constant \cite{Hawking, Duff, Wu,Duncan,Bousso:2000xa}.
The 3-form can gain a mass via the St\"uckelberg mechanism by
``eating''  a 2-form and as a consequence 
a massive  3-form has one physical scalar degree of freedom.

In $N=1$ supersymmetric theories
the massless 3-form resides in a 
supermultiplet together with  two scalar and two fermionic physical degrees
of freedom \cite{Gates:1980ay,Binetruy:1996xw,Ovrut:1997ur} and
Poincar\'e duality relates this multiplet to a non-minimal scalar multiplet 
discussed in \cite{Gates:1980az,Corresp}.\footnote{%
Since 3-forms can also play the role of the Yang-Mills Chern-Simons
term, the 3-form multiplet
has been used in the description of  gauge anomalies in supersymmetric
theories \cite{Girardi:1985hf,Girardi:1998ju,Binetruy:2000zx}.}
A massive 3-form multiplet gains its mass  by ``eating'' a linear multiplet
which contains a 2-form as one of its components.
The linear multiplet also has  two scalar and two fermionic 
physical degrees of freedom so that a massive 3-form multiplet
altogether has four scalar and four
fermionic degrees of freedom. Its Poincare dual multiplet is closely related
to the complex linear multiplet studied in \cite{comments,nonmsca}.
The massive 3-form multiplet appears in the effective description of
gaugino condensation and was used to describe a (massive) 
pseudo-scalar glueball \cite{FGS1,FGS2,noteSYM}.

In this paper we systematically study the massless and massive 3-form
multiplet, 
their renormalizable and non-renormalizable actions and their
Poincar\'e dual descriptions. 
We  pay particular attention to the issue
of boundary terms which in the non-supersymmetric case were
discussed in  \cite{Duff, Wu,Duncan,Bousso:2000xa} while, as far as we
know,
the supersymmetric version does not exist in the literature so far.

For the case of a non-renormalizable sigma model 
we find that the scalar fields 
of the 3-form multiplets 
span a K\"ahler manifold. 
The dual scalars have the same geometry but with the K\"ahler metric
expressed in terms of 
the Legendre transform of the original K\"ahler potential.
A massive 3-form multiplet features an additional real scalar whose 
sigma model metric is related to the  mass matrix of
the 3-forms. 
In the dual action the massive 3-form is replaced by another scalar
and the resulting geometry is the product of two K\"ahler manifolds.

This work is organized as follows. In section~\ref{sec:two1}
we introduce the 3-form multiplet and give its field strength
as well as its gauge and supersymmetry transformations.
In section~\ref{RenormMassless} we give the renormalizable kinetic
action and discuss the necessary supersymmetric boundary terms.
In section~\ref{RenormMassive} we introduce mass and Fayet-Iliopoulos terms 
and compute the component action, while the modification of the mass spectrum
in the presence of a superpotential is discussed in \ref{sec:superpot}. 
The dual actions in the massless and massive
case are derived in sections~\ref{ch_dual_renorm_massl}
and \ref{ch_dual_renorm_massive} respectively.
Section~\ref{sec:three} generalizes the analysis to non-renormalizable
couplings. Here we confine our attention to the bosonic terms
in the action and focus on the appearing target space geometries.
In section~\ref{ch_coupling} we study the coupling of massive 3-form
multiplets to chiral multiplets.
We conclude in section~\ref{sec:five} and some additional material is
assembled in six appendices.
Appendix~\ref{conventions} summarizes our conventions while
Appendix~\ref{multiplets} recalls the different $N=1$ supermultiplets.
In Appendix \ref{app_3form} we review
the issues of boundary terms and boundary conditions for the 3-form action in
non-supersymmetric theories and its connection with the cosmological
constant.
Appendix~\ref{elim_aux} provides some generic formulas for eliminating
auxiliary fields which we frequently use in the main text.  
Appendix~\ref{app:real} discusses the dual action for the special case
of K\"ahler
potentials with a shift symmetry.
Finally,
Appendix~\ref{App_Legendre} gives a brief introduction to 
the Legendre transformation.

\section{The 3-form multiplet}

\subsection{Components, field strength and gauge transformation}
\label{sec:two1}

The  superfield $U$ which contains  a 3-form $C_{npq}$
can be constructed from a vector multiplet
with the vector field defined as\footnote{Our conventions  are summarized in Appendix \ref{conventions}
while basic facts
  about the various supermultiplets used in this
  paper are recalled in Appendix~\ref{multiplets}.}
\begin{equation} \label{v_m}
v_{m} = \tfrac{1}{6}\, \varepsilon_{mnpq}C^{npq}\ , \qquad m,n,p,q = 0,\ldots,3\ .
\end{equation}
It reads\footnote{Compared to the usual normalization we rescaled $U$
  in \eqref{def_U} and \eqref{defS} by a factor of $16$ for later
  convenience.} \cite{FGS1,Ovrut:1997ur}
\begin{equation}\bal \label{def_U}
	U  = \bar U =\  &B + i\theta\chi - i\bar{\theta}\bar{\chi} + \thz M^* + \thq M
	+\tfrac{1}{3}\theta \sigma^{m}\bar{\theta}\veps_{mnpq}C^{npq}\\
	&+\thz\bar{\theta} \big(\sqrt{2}\bar{\lambda} + \tfrac{1}{2}\bar{\sigma}^{m}\del_{m}\chi \big) + \thq\theta \big(\sqrt{2}\lambda - \tfrac{1}{2}\sigma^{m}\del_{m}\bar{\chi} \big)
	+\thvoll \big(D - \tfrac{1}{4}\Box B \big)\ ,
\eal\end{equation}
where  $B$ and $D$ are real scalars, $M$ is a complex scalar, and $\chi$ and $\lambda$ are Weyl spinors. Like the vector multiplet, $U$ carries 8 fermionic ($\chi$, $\lambda$) and 8 bosonic ($B, D, M$ and $C^{npq}$)  degrees of freedom off-shell.
The difference between the 3-form multiplet and the vector multiplet is
only visible in the definitions of their field strengths. A vector as a $1$-form has a $2$-form field strength, with another $2$-form as its dual. The field strength of the $3$-form on the other hand is a $4$-form
\begin{equation}\label{defH3}
 H_{mnpq} = 4\del_{[m}C_{npq]} =  \del_{m}C_{npq}-\del_{n}C_{pqm}+\del_{p}C_{qmn}-\del_{q}C_{mnp} \ ,
\end{equation}
with a  $0$-form $H$ as its dual,
\begin{equation} \label{defH}
	H = \tfrac{1}{4!}\varepsilon^{mnpq}H_{mnpq}\ ,  \qquad
	H_{mnpq} = -\varepsilon_{mnpq}H\ .
\end{equation}
For this reason a field strength for the 3-form multiplet cannot be
constructed like the vector multiplet's $W_\alpha = -\frac{1}{4} \bar
D^2 D_\alpha V$ defined in \eqref{vector_fieldstrength}. Instead it is
defined by \cite{Gates:1980ay,Buchbinder:1988tj,Binetruy:1996xw,Ovrut:1997ur}
\begin{equation}\label{defS}
S = - \tfrac{1}{4} \bar{D}^{2} U\ ,
\end{equation}
which implies that $S$ is chiral
\begin{equation} \label{constchir}
	\bar{D}_{\dot{\alpha}} S = 0\ .
\end{equation}
Its expansion in component fields reads
\begin{equation}\bal \label{S_components}
	S =\  M +i\theta\sigma^{m}\bar{\theta}\del_{m}M + \tfrac{1}{4}\thvoll\Box M
	+\sqrt{2}\theta \lambda+\tfrac{i}{\sqrt{2}} \theta \theta \bar{\theta} \bar{\sigma}^{m} \del_{m} \lambda
	+\theta\theta (D + iH)\ . 
\end{aligned}\end{equation}
Since $U$ is real $S$ is not a
generic chiral superfield.\footnote{Recall that every chiral superfield can be
  expressed as $\Phi = \bar{D}^{2} F$ with  $F$ being complex.} Indeed, the imaginary part of the
$\thz$-component of $S$ is the dual field strength $H$, which, being a
total divergence, is not an unconstrained scalar field.

From its definition (\ref{defS}) we see that $S$ is invariant under the gauge transformation 
\begin{equation}\label{gaugeU}
 U \rightarrow\; U - L\ ,  
\end{equation}
where $L$ is a linear multiplet obeying $D^{2}L=\bar{D}^{2}L=0$.
In particular,
$L$ contains a real scalar $E$, a Weyl fermion $\eta$
and the field strength  of a two-form
$\del_{[n}B_{pq]}$ as component fields
(see Appendix~\ref{ch_linearmultiplet} for further details).
Using \eqref{linearmulti} one infers  the gauge transformation
of the component fields to be
\begin{equation}\label{gauge_U_comp}
\bal
	B &\rightarrow B-E\ ,\quad \chi \rightarrow \chi-\eta\ ,
\quad M \rightarrow M\ , \\\ 
	C_{npq} &\rightarrow C_{npq}-\del_{[n}B_{pq]}\ ,\quad \lambda
        \rightarrow \lambda\ , \quad D \rightarrow D \ ,
\eal 
\end{equation}
which renders $B$ and $\chi$ gauge degrees of freedom. 
Thus the massless 3-form multiplet $U$ has
four scalar and four fermionic off-shell degrees of freedom, which occur in the field strength multiplet \eqref{S_components}.\footnote{Note that
since $S$ is a chiral superfield it naturally carries mass dimension
1. From \eqref{defS} and the fact that $D_\alpha$ 
has dimension $1/2$ it follows that $U$  has mass dimension 0. 
$\theta_\alpha$  has dimension $-1/2$ and thus 
\eqref{def_U} implies dimension 0 for $B$ and 1/2 for $\chi$, i.e.\
both fields have non-canonical mass dimensions.
Therefore, when they enter the massive 3-form action as will be 
described in sec.~\ref{RenormMassive}, they have to be rescaled in order to get kinetic terms of the standard form.\label{ftmass}}



The supersymmetry transformations of the 3-form multiplet are given 
by \cite{Binetruy:1996xw,Ovrut:1997ur}
\begin{equation} \label{SUSY_3-form}
\bal 
 \delta_\xi M &= \sqrt{2}\xi\lambda\ , \\
 \delta_\xi \lambda &= \sqrt{2}i\sigma^m\bar\xi \del_m M + \sqrt{2} \xi(D+iH)\ ,\\
 \delta_\xi C_{npq} &= \veps_{mnpq}\xi\l(\tfrac{1}{\sqrt{2}} \sigma^m\bar\lambda + \sigma^{ml}\del_l\chi \r) + \tn{h.c.}\ ,\\
 \delta_\xi B &= i\xi\chi - i\bar\xi\bar\chi\ ,\\
 \delta_\xi \chi &= -2i \xi M^* + \sigma^m \bar\xi \l(-\tfrac{i}{3}\veps_{mnpq}C^{npq} + \del_m B\r)\ .
\eal
\end{equation}
Note that the gauge invariant components of $S$ transform among
themselves as 
in an ordinary chiral multiplet. Also note that the second term in the supersymmetry variations of the 3-form (which does not appear in the references 
\cite{Binetruy:1996xw,Ovrut:1997ur}) constitutes only a gauge transformation since it can be written as
\begin{equation}
 \veps_{mnpq} \xi\sigma^{ml}\del_l\chi = \tfrac{3}{2} \del_{[n}\veps_{pq]ml} \xi\sigma^{ml}\chi\ .
\end{equation}

\subsection{Renormalizable action of the massless 3-form multiplet} \label{RenormMassless}

We are now prepared to construct the gauge invariant kinetic 
action for $N_3$ massless 3-form multiplets 
$U^a,\ a=1, \ldots ,N_3$ with field strengths $S^a$. 
Since the $S^a$ are chiral, the kinetic action is given by the standard 
expression\cite{Corresp}
\begin{equation}\bal \label{S_kin}
S_3 = \ivoll \delta_{a\bb} S^a \bar S^{\bb} 
  = \ix \delta_{a\bb}\big(-\del_m M^a \,\del^m M^{*\bb} - i \lambda^a \sigma^m \del_m \bar\lambda^{\bb} + D^a D^{\bb} + H^a H^{\bb} \big)\ ,
\eal \end{equation}
where $D^{\bb} = D^b$ and $H^{\bb} = H^b$ since both are
real.\footnote{To obtain the component action we used partial integration 
for the fields $M^a$ and $\lambda^a$ which might seem questionable 
since we cannot assume that boundary terms involving the
  3-form field strengths $H^a$ vanish  and
  the latter transform into these fields under supersymmetry. The issue of
the boundary terms is discussed in more detail in the following and in
Appendix~\ref{app_3form}.
However, for now we can note that boundary terms do not affect the equations of motion and therefore it is
legitimate to drop them here.}
The  $D^a$ are auxiliary fields which vanish by their equations of motion 
$D^a=0$. 
The action \eqref{S_kin} contains the correct kinetic term for the 3-form
\begin{equation} \label{H2}
	H^{a}H^{a} = -\tfrac{1}{24}H^a_{mnpq}H^{amnpq}\ .
\end{equation}
However, as we discuss in more detail in Appendix \ref{app_3form}, 
one has to add a boundary term to the action \eqref{S_kin} 
in order to impose 
the gauge invariant variational 
boundary condition $\delta H^a\vert_{\del\mathcal M} = 0$ instead of the gauge variant $\delta C^a_{npq}\vert_{\del\mathcal M} = 0$ required by \eqref{S_kin}  \cite{Wu,Duncan}.
(Here $\mathcal M$ is the integration volume.)
In a supersymmetric theory the boundary condition should in addition
be supersymmetry invariant and thus we demand
\begin{equation} \label{VC_susy}
 \delta S^a\Big\rvert_{\del \mathcal M} = 0\ , \qquad D_\alpha (\delta S^a)\Big\rvert_{\del \mathcal M} = 0
\end{equation}
rather than $\delta U^a\rvert_{\del \mathcal M} = 0, \ D_\alpha (\delta U^a)\rvert_{\del \mathcal M} = 0$.
This can be achieved by adding to \eqref{S_kin} 
the boundary term
\begin{equation}\label{boundary_terms}
\bal \mathcal B \ &= \ \tfrac{1}{4} \ivoll D^\alpha\big(S^a D_\alpha U^a - (D_\alpha S^a) U^a\big) + \hc \\
      &= \   \Re \ivoll D^\alpha\l(S^a D_\alpha U^a\r) - \tfrac{1}{2} \Re \ivoll D^2 \l(S^a U^a\r)\ , \eal 
\end{equation}
where the first term in the second line of \eqref{boundary_terms} contains the 
boundary term for the 3-forms
\begin{equation} \label{bt_3form}
  - \tfrac{1}{3} \ix  \del_m \big(H^a \veps^{mnpq}  C^a_{npq}\big)\ .
\end{equation}
One can easily check that the variation of the action
\begin{equation}  \label{S_massl}
 S_3^\prime = S_3 + \mathcal B
\end{equation}
with respect to $U$ is given by
\begin{equation}\label{deltaS}
 \delta S_3^\prime = \tfrac{1}{4}\ivoll \Big[\!-(D^2 S^a) \delta U^a +  D^\alpha \big((\delta S^a) D_\alpha U^a - D_\alpha(\delta S^a) U^a \big) + \hc \Big]\ .
\end{equation}
Applying 
the  constraint \eqref{VC_susy} to the variation \eqref{deltaS}
leads to 
the superfield equations of motion 
\begin{equation} \label{eom_S}
 D^2 S^a + \bar D^2 \bar S^a = 0\ .
\end{equation}
They include the equations of motion for $C^a_{npq}$ 
\begin{equation}
\veps^{mnpq} \del_m H^a =0\ ,
\end{equation}
which imply that the 3-form field strengths are constants,
i.e.\ $H^a = c^a$ with $c^a \in \mathbb R$. Thus the action \eqref{S_kin} describes $2N_3$ bosonic and $2N_3$ fermionic degrees of freedom on-shell.
Due to the presence of the boundary term \eqref{bt_3form} one can check 
that the 3-forms contribute a \emph{positive} constant potential
\cite{Hawking, Duff, Wu, Duncan, Bousso:2000xa}
\begin{equation}
 V = c^a c^a\ ,
\end{equation}
corresponding to a positive correction of the bare cosmological constant $\Lambda_0$\footnote{Without the boundary term the contribution would be negative 
(for a detailed discussion see Appendix \ref{app_3form}).}
\begin{equation}
 \Lambda = \Lambda_0 + 8\pi G c^a c^a\ .
\end{equation}
The supersymmetry variation of $\lambda$ given in \eqref{SUSY_3-form} shows that it transforms inhomogeneously for $H^a=c^a\neq 0$ and thus
supersymmetry is spontaneously broken with $\lambda$ being 
the Goldstone fermion.

\subsection{Renormalizable action of the massive 3-form multiplet} \label{RenormMassive}

Let us now consider a massive 3-form multiplet by
adding a gauge invariant mass term 
to the action \eqref{S_kin} with the help of the St\"uckelberg mechanism. 
One introduces $N_3$ additional linear multiplets $L^{\prime a}$ with the transformation law
\begin{equation} \label{gaugeL}
	L^{\prime a} \to L^{\prime a}-L^{a},
\end{equation}
where the transformation parameters $L^{a}$ are also linear
superfields. Due to \eqref{gaugeU} the combination $U^a-L^{\prime a}$ is
gauge invariant
so that one can add to the action the mass term
\cite{noteSYM}
\begin{equation} \label{renorm_massterm}
	S_{\text{mass}} = \ivoll \big(- \tfrac{1}{2}m_{ab}^{2}(U^{a}-L^{\prime a})(U^{b}-L^{\prime b})+\xi_{a}(U^a-L^{\prime a}) \big)\ ,
\end{equation}
where $m_{ab} = m_{ba}$ is a symmetric mass matrix  and the $\xi_a$ parametrize
possible Fayet-Iliopoulos terms.\footnote{Note that cubic terms in $U$ are non-renormalizable.}
The additional degrees of freedom introduced by the $L^{\prime a}$ can
be absorbed into the $U^a$ by fixing a gauge. In the following we work
in the ``unitary''
gauge $L^{\prime a}=0$.
Furthermore, for simplicity 
we assume that there are no massless modes 
and  thus $m_{ab}$ is invertible.\footnote{The action for 
massless modes was already given in section~\ref{RenormMassless}.}
In the gauge $L^{\prime a}= 0$ we have the action\footnote{
The boundary terms \eqref{boundary_terms} should be added to the massive action as well. Since we are not going to eliminate the massive 3-forms here, this is however not relevant for our purposes.}
\begin{equation}\bal \label{actionmassiv}
	S_3 = &\ivoll \big(\delta_{a\bb}S^{a}\bar S^{\bb} - \tfrac{1}{2} m^{2}_{ab} U^a U^b + \xi_{a} U^{a} \big)	\\
	= & \int\! d^4x \Big(-\del_{m}M^{a}\,\del^{m}M^{a*} -i\lambda^{a}\sigma^{m}\del_{m}\bar{\lambda}^{a}+D^{a}D^{a} + H^{a}H^{a} \\
	& \qquad-\tfrac{1}{2}m^{2}_{ab}\big(i\chi^{a}\sigma^{m}\del_{m}\bar{\chi}^b
	-\sqrt{2}i\chi^a\lambda^b + \sqrt{2}i\bar{\chi}^a\bar{\lambda}^b
	+ 2M^{a}M^{b*} \\
	& \qquad + 2B^aD^b - \tfrac{1}{2}B^{a}\Box B^b +\tfrac{1}{3}C^a_{npq}C^{bnpq} \big)
	+ \xi_a (D^{a}-\tfrac{1}{4}\Box B^{a}) \Big)\ .
\eal\end{equation}
The auxiliary fields $D^a$ can be eliminated by their equations of motion
\begin{equation} \label{eomD}
	2\delta_{ab}D^{b} - m^{2}_{ab}B^b + \xi_{a} = 0\ .
\end{equation}
This is done most conveniently by  ``completing the square'' as described in Appendix \ref{elim_aux},
leading to the on-shell action
\begin{equation} \label{S_renorm_massive}
\bal S_3 = \ix \Big(&-\del_{m}M^{a}\,\del^{m}M^{a*}-i\lambda^{a}\sigma^{m}\del_{m}\bar{\lambda}^{a} + H^{a}H^{a}  \\
  & -m^{2}_{ab}\big(\tfrac{i}{2}\chi^{a}\sigma^{m}\del_{m}\bar{\chi}^b -\tfrac{i}{\sqrt{2}}\chi^{a}\lambda^b + \tfrac{i}{\sqrt{2}}\bar{\chi}^{a}\bar{\lambda}^b + M^a M^{b*} \\
  &  + \tfrac{1}{4}\del^m B^{a}\del_m B^b
  +\tfrac{1}{6}C^a_{npq}C^{bnpq} \big) -\tfrac{1}{4}(m^2_{ab}B^b -
  \xi_a)(m^2_{ac}B^c - \xi_a) \Big)\ . \eal
\end{equation}

As we already mentioned in footnote~\ref{ftmass}, $B^a$ and $\chi^a$ do
not have standard mass dimensions. In the previous section this was
irrelevant as both fields dropped out as gauge degrees of freedom.
For a massive 3-form multiplet however, they become physical and in
order for their kinetic terms to have the canonical form we need the following field redefinitions
\begin{equation} \label{B_chi_redef}
 \bal B^{\prime a} := \tfrac{1}{2}\big(\delta^{ab}m_{bc}B^c - m^{-1ab}\xi_b\big) \ ,\qquad
    \chi^{\prime a} := -\tfrac{i}{\sqrt{2}} \delta^{ab}m_{bc} \chi^c\ . \eal
\end{equation}
Note that we also shifted $B$ by a constant proportional to the
FI-parameter $\xi$. From  \eqref{S_renorm_massive} we see that as long as
$m^2_{ac}$ has maximal rank $\xi$ merely induces a  vacuum expectation
value for $B$ but does not break supersymmetry. However, whenever 
$m^2_{ac}$ has a zero eigenvalue and the corresponding $\xi$ is
non-zero,
supersymmetry is spontaneously broken. In the following we assume
$m^2_{ac}$ to have  maximal rank and perform the field redefinition 
given in \eqref{B_chi_redef}.
 Then the on-shell action is independent of $\xi$ and reads
\begin{equation}\bal \label{S_renorm_mass_onshell}
	S_3 = \int\! d^4x \big(&-\del^{m}M^{a}\,\del_{m}M^{a*} - m^{2}_{ab}M^{a}M^{b*}
	-\del^{m}B^{\prime a}\del_{m}B^{\prime a} - m^{2}_{ab}B^{\prime a}B^{\prime b} \\
	&-i\lambda^{a}\sigma^{m}\del_{m}\bar{\lambda}^{a} -i\chi^{\prime a}\sigma^{m}\del_{m}\bar{\chi}^{\prime a}
	- m_{ab}\chi^{\prime a}\lambda^b - m_{ab}\bar{\chi}^{\prime a}\bar{\lambda}^b \\
	& + H^{a}H^{a} -\tfrac{1}{6}m^{2}_{ab}C^a_{npq}C^{bnpq} \big)\
        .
\eal\end{equation}
We see that the fermions $\lambda^a$ and $\chi^{\prime a}$ form $N_3$
massive Dirac-spinors corresponding to $4N_3$ fermionic degrees of freedom. 
The $N_3$ massive 3-forms now contribute one bosonic on-shell degree of freedom
each, because their equations of motion
\begin{equation}
  -\delta_{ab}\, \veps^{mnpq}\del_m H^b  =  m^2_{ab}C^{bnpq}\ ,
\end{equation}
which imply
\begin{equation}
 \del^n C^b_{npq} = 0, \qquad \delta_{ab} \Box C^b_{npq} = m^2_{ab}C^b_{npq}\ ,
\end{equation}
remove $3N_3$ of the $4N_3$ off-shell degrees of freedom.
Together with $N_3$ complex  scalars $M^a$ and the 
$N_3$ real scalars $B^{\prime a}$ we thus also have $4N_3$ massive bosonic on-shell
degrees of freedom. 

\subsection{Including a superpotential}\label{sec:superpot}

Since the $S^a$ are chiral superfields, one may also add a superpotential to the action. This is an alternative way to introduce masses for the components of the 3-form multiplet, although the 3-form itself cannot gain a mass in this way as we will see below. A superpotential can also lead to spontaneous supersymmetry breaking as in the case of ordinary chiral multiplets. For simplicity we drop the mass term of the previous section and start with the action
\begin{equation}\label{S_superpot}
\bal S_3 &= \ix \bigg[\idtdtq S^a \bar S^a  + \idt W(S) + \idtq W^*(\bar S) \bigg]\\
  &= \ix \Big(-\del_m M^a \,\del^m M^{a*} - i \lambda^a \sigma^m \del_m \bar\lambda^{a} + D^a D^a + H^a H^a \\
  & \qquad + W_a(D^a\!+\!iH^a) + W_a^*(D^a\!-\!iH^a) - \tfrac12 W_{ab}\lambda^a\lambda^b - \tfrac12 W_{ab}^*\bar\lambda^a\bar\lambda^b \Big)\ ,\eal
\end{equation}
where
\begin{equation}
 W_a(M) := \der{W}{S^a}\bigg|_{\theta=\btheta=0}, \quad W_{ab}(M) := \dder{W}{S^a}{S^b}\bigg|_{\theta=\btheta=0}\ .
\end{equation}
The auxiliary fields $D^a$ can be easily eliminated from \eqref{S_superpot}, creating a contribution to the scalar potential
\begin{equation}
 \mathcal V_D =  \big(\Re (W_a) \big)^2\ .
\end{equation}
However, also the massless 3-forms have to be eliminated in order to find the effective scalar potential. Their equations of motion read
\begin{equation}
\veps^{mnpq} \del_m \big(H_a - \Im (W_a)\big) = 0\ ,
\end{equation}
where $H_a = \delta_{ab}H^b$, and they have the solution $H_a = \Im(W_a) + c_a$ with $c_a \in \mathbb R$.
In order to impose gauge invariant boundary conditions one again has to add appropriate boundary terms to the action \eqref{S_superpot}.
We will not do this in a supersymmetric way here, but only give the correct boundary term for the 3-forms which reads (cf.~\eqref{bt_3form})
\begin{equation} \label{bt_3form_superpot}
  \mathcal B_3 = - \tfrac{1}{3} \ix  \del_m \Big(\big(H_a - \Im(W_a)\big) \veps^{mnpq}  C^a_{npq}\Big)\ .
\end{equation}
When this term is added to the action, the contribution of the 3-forms to the scalar potential is found to be
\begin{equation}
 \mathcal V_3 = H_a H_a\Big\rvert_{H_a\; =\; \Im(W_a)\; +\; c_a}\ .
\end{equation}
Thus the effective scalar potential is given by
\begin{equation}\label{scalar_pot}
 \mathcal V = W_a W_a^* + 2c_a \Im(W_a) + c_a c_a = (W_a + ic_a)(W_a^*-ic_a)\ .
\end{equation}
This coincides with the result for ordinary chiral multiplets with the modified superpotential\cite{Binetruy:1996xw,noteSYM}
\begin{equation}
 \tilde W(S) = W(S) + ic_a S^a\ .
\end{equation}
Thus the analysis of spontaneous supersymmetry breaking and the mass spectrum can be performed in exactly the same way as for the well known O'Raifeartaigh models\cite{O'Raifeartaigh:1975pr,WB}. We find that a superpotential can create masses for the scalars $M^a$ and Weyl spinors $\lambda^a$ but the 3-forms
 remain massless.

One may also consider an action 
\begin{equation} \label{S_superpot_mass}
 \bal S_3 &= \ix \bigg[\idtdtq \Big(S^a \bar S^a - \tfrac{1}{2} m^{2}_{ab} U^a U^b + \xi_a U^a\Big)  + \idt W(S) + \idtq W^*(\bar S) \bigg] \eal\\
\end{equation}
that contains both a superpotential and mass and FI terms for the $U^a$ as done in \cite{FGS1,FGS2,noteSYM} for the low energy effective description of $N=1$ SYM theory. The 3-forms are then dynamical field variables that cannot be eliminated from the action. The only auxiliary fields of the theory are the $D^a$ that couple both to the $B^a$ as in \eqref{actionmassiv} and to the real part of $W_a$ as in \eqref{S_superpot}. After their elimination and rescaling of $B$ and $\chi$ as in \eqref{B_chi_redef} one obtains the action (dropping the prime on $B^a$ and $\lambda^a$)
\begin{equation}\label{S_superpot_mass_comp}
\bal S_3 =  \ix \Big(&-\del_m M^a \,\del^m M^{a*} - i \lambda^a \sigma^m \del_m \bar\lambda^{a} + H^a H^a   -\del^m B^a \del_m B^{a} -i\chi^a \sigma^m\del_m\bar{\chi}^a\\
&- m_{ab}\l(\chi^{a}\lambda^b + \bar{\chi}^{a}\bar{\lambda}^b\r) - \tfrac{1}{2}W_{ab}\lambda^a\lambda^b -  \tfrac12 W_{ab}^*\bar\lambda^a\bar\lambda^b - m^{2}_{ab}M^{a}M^{b*} \\
& -\tfrac{1}{6}m^{2}_{ab}C^a_{npq}C^{bnpq} - 2\Im(W_a)H^a 
-\l(m_{ab}B^{b} - \Re(W_a)\r)^2\Big)\ .
   \eal
\end{equation}
The equation of motion for $C^a_{npq}$ reads
\begin{equation}\label{eom_C}
  -\veps^{mnpq}\del_m \big(H_a - \Im (W_a)\big)  =  m^2_{ab}C^{bnpq}\ .
\end{equation}
Since the massive 3-forms carry one physical degree of freedom each, they can be represented by the real scalars
\begin{equation}
 \phi^a := m^{-1ab}\big(H_b - \Im (W_b)\big)\ .
\end{equation}
It follows from \eqref{eom_C} that this scalar satisfies the equation of motion
\begin{equation}\label{eom_phi}
 \delta_{ab} \Box \phi^b = m^2_{ab}\phi^b  + m_{ab} \Im (W_b)\ ,
\end{equation}
while the equations of motion for $B^a$ and $M^a$ are
\begin{equation}\label{eom_B_M}
 \bal \delta_{ab} \Box B^b &= m^2_{ab}B^b - m_{ab} \Re(W_b)\ ,\\  \delta_{ab} \Box M^b &= m^2_{ab}M^b + W^*_{ab}W_b + W^*_{ab}m_{bc} \big(\!-\!B^c+i\phi^c\big)\ . \eal 
\end{equation}
These equations call for the definition of the $N_3$ complex scalar fields
\begin{equation}
 N^a := -B^a + i\phi^a\ .
\end{equation}
In the theory dual to \eqref{S_superpot_mass_comp}  where the 3-forms are replaced by the scalars $\phi^a$, the scalar potential is given by
\begin{equation}\label{V_dual}
 \mathcal V\ =\ m^2_{ab}M^aM^{b*} + (m_{ab}N^b + W_a)(m_{ac}N^{c*} + W_a^*)\ ,
\end{equation}
as can be easily seen from the equations of motion for $M^a$ and $N^a$. As $\mathcal V$ vanishes for $M^a = 0$, $N^a = - m^{-1ab}W_b\vert_{M=0}$, supersymmetry again remains unbroken by virtue of the non-singular mass matrix $m_{ab}$. Moreover we find that $\langle S^a \rangle = 0$.

As linear terms in $W$ only induce a shift in the VEVs of the $N^a$, one can absorb them into the $N^a$ by redefining $N^a \to \langle N^a \rangle + N^a$. Thus we can assume without loss of generality that the superpotential is of the form
\begin{equation}\label{form_superpot}
 W(S) = \tfrac12 \mu_{ab} S^a S^b + \mathcal O(S^3)
\end{equation}
with a symmetric matrix $\mu$. Then the mass terms of the scalar potential \eqref{V_dual} can be identified as
\begin{equation} \label{mass_terms}
 \mathcal V = \big(M^\dagger \  N^\dagger\big) \begin{pmatrix} \mu^*\mu + m^2 & \mu^*m \\ m\mu & m^2 \end{pmatrix} \begin{pmatrix} M \\ N \end{pmatrix} + \tn{higher order terms} \ .
\end{equation}

The quadratic mass matrix for the fermions $\lambda^a$ and $\chi^a$ can be directly read off from the action  \eqref{S_superpot_mass} and is given by
\begin{equation}
 m^2_{\lambda,\chi} =  \begin{pmatrix} \mu^* & m\\ m & 0 \end{pmatrix} \begin{pmatrix} \mu & m\\ m & 0 \end{pmatrix} = \begin{pmatrix} \mu^*\mu + m^2 & \mu^*m \\ m\mu & m^2 \end{pmatrix}\ .
\end{equation}
It is identical with the mass matrix for the scalars $M^a$ and $N^a$ found in \eqref{mass_terms}.
In the one dimensional case ($N_3=1$) the mass eigenvalues are given by
\begin{equation}
 m_\pm = \l|\tfrac12 |\mu| \pm \sqrt{\tfrac14 |\mu|^2 + m^2}\r|\ .
\end{equation}
Note that this coincides (for real $\mu$) with the result given in \cite{noteSYM} with the correspondence $\mu = m_{11},\ m=m_{12}$. Obviously, 
the four bosonic and fermionic mass eigenstates organize into two chiral supermultiplets with masses $m_+$  and $m_-$ respectively (as this is the only $N=1$ supermultiplet which contains only particles of spin 0 and 1/2).

\subsection{Dualization of the massless action} \label{ch_dual_renorm_massl}

It is possible to reproduce a physical action from a 
`first order action' by introducing additional fields with algebraic 
equations of motion \cite{forder}. Alternatively the original fields can be eliminated from the first order action, giving rise to a dual action. 
There is a one-to-one map between the on-shell degrees of freedom of action and dual action which is defined by the Euler-Lagrange equations of the first order action.

For the case at hand the massless action (\ref{S_massl}) can be reproduced from the first order action \cite{Corresp}
\begin{equation} \label{S_first_renorm}
	S_\tn{first} = \ivoll (-\delta^{a\bb}F_a \bar F_\bb + F_a S^a + \bar{F}_\ba \bar{S}^\ba) + \mathcal{B}_\tn{first}\ ,
\end{equation}
including the boundary term
\begin{equation} \label{B_first}
  \mathcal{B}_\tn{first} = \tfrac{1}{4} \ivoll \big(\bar D_\dalpha \l(F_a \bar D^\dalpha U^a - \bar D^\dalpha F_a  U^a\r) + \hc \big)\ .
\end{equation}
Here the $F_a$ are unconstrained superfields with a component field expansion 
\begin{equation}\begin{aligned} \label{defF}
	F_a =\ &f_a + \theta\psi_a + \sqrt{2}\bar{\theta}\bar{\varphi}_a+\thz h_a + \thq n_a+\thsigmath w_{am} \\
&+ \thz\bar{\theta}\bar{\vartheta}_a + \thq\theta(\zeta_a - \tfrac{i}{\sqrt{2}}\sigma^{m}\del_{m}\bar{\varphi}_a) + \thvoll(d_a - \tfrac{1}{4}\Box f_a - \tfrac{i}{2}\del_{m}w_a^{m}),
\end{aligned}\end{equation}
where $f_a,\ h_a,\ n_a$ and $d_a$  are complex scalars, $w_{am}$ is a complex vector and $\psi_a,\ \varphi_a, \vartheta_a$ and $\zeta_a$ are Weyl spinors. 
The $F_a$ can be eliminated from (\ref{S_first_renorm})
by inserting their equations of motion
\begin{equation} \label{eom_F}
	\delta^{a\bb}\bar F_\bb = S^a\ ,
\end{equation}
reproducing  the action (\ref{S_massl}).

To find the dual action one inserts
(\ref{defS}) to obtain 
\begin{equation}\begin{aligned} \label{S_first2}
	S_\text{first} &= \ivoll \big(- \tfrac{1}{2} \delta^{a\bb}F_a \bar F_\bb - \tfrac{1}{4} F_a  \bar D^2 U^a + \tfrac{1}{4} \bar D_\dalpha \l(F_a \bar D^\dalpha U^a - \bar D^\dalpha F_a  U^a\r) + \hc \big)\\
	&= \ivoll \big(- \delta^{a\bb}F_a \bar F_\bb - \tfrac{1}{4} \l(\bar{D}^2 F_a + D^{2}\bar F_a\r)U^a\big)\ ,
\end{aligned}\end{equation}
where the Leibniz rule for the covariant superspace derivatives was applied. 
Note that due to the  boundary term the first order action 
has a simple form whose  variation with respect to $U^a$ yields immediately 
(and without dropping any boundary term) a constraint for~$F_a$
\begin{equation}\label{eomU_massl}
\bal
    0\ &= \ -\tfrac{1}{4}\l(\bar{D}^{2} F_a + D^{2} \bar F_a\r) \\
      &= \ n_a + n^*_a + \theta\zeta_a + \bar{\theta}\bar{\zeta}_a + \thz d_a + \thq d_a^{*}
	+ i\thsigmath\del_{m}(n_a-n^{*}_a) \\
	&\quad +\tfrac{i}{2} \thz\bar{\theta}\bar{\sigma}^{m}\del_{m}\zeta_a + \tfrac{i}{2} \thq\theta\sigma^{m}\del_{m}\bar{\zeta}_a + \tfrac{1}{4}\thvoll\Box(n_a+n_a^*)\ .
\eal
\end{equation}
As usual, Poincar\'e duality has exchanged the equation of motion with the constraint (cf. \eqref{eom_S}). We will see below that the condition (\ref{eomU_massl}) is special for 
the massless case in that it reduces the number of degrees of freedom in $F_a$
while in the massive case the $F_a$ remain unconstrained superfields.
As implied by (\ref{eomU_massl}) $\zeta_a$ and $d_a$ vanish
while the $n_a$ become purely imaginary constants,
\begin{equation}
 \zeta_a = 0\ , \qquad d_a = 0\ , \qquad n_a = i\cc_a\ , \quad \tn{with} \quad \cc_a \in \mathbb{R}\ .
\end{equation}
Therefore $F_a$ takes the form
\begin{equation} \begin{aligned} \label{F_constr}
 F_a =\ &f_a + \theta\psi_a+\sqrt{2}\bar{\theta}\bar{\varphi}_a + \thz h_a + i\thq \cc_a+\thsigmath w_{am}
	+\thz\bar{\theta}\bar{\vartheta}_a-\tfrac{i}{\sqrt{2}}\thq\theta\sigma^{m}\del_{m}\bar{\varphi}_a \\
	& + \thvoll (-\tfrac{i}{2}\del_{m}w_a^{m}-\tfrac{1}{4}\Box f_a)\ ,
\eal \end{equation}
containing 12 bosonic and 12 fermionic off-shell degrees of freedom. 
Using \eqref{eomU_massl} and \eqref{F_constr} we obtain as the dual 
component action
\begin{equation}\begin{aligned} \label{S_dual_massless1}
S_{\text{dual}} &= \ivoll (-\delta^{a\bb}F_a \bar F_\bb) \\
	    &= \ix \Big(f_a^{*}(\tfrac{i}{2}\del_{m}w_a^{m} + \tfrac{1}{4}\Box f_a)+\tfrac{1}{2}\psi_a\vartheta_a
-\tfrac{i}{2}\varphi_a\sigma^{m}\del_{m}\bar{\varphi}_a \\
 &\quad\qquad -\tfrac{1}{2}h_a h_a^{*}-\tfrac{1}{2}\cc_a \cc_a +
 \tfrac{1}{4}w_{am}^{*}w_a^{m} + \tn{h.c.} \Big)\ .
\end{aligned}\end{equation}
After eliminating the auxiliary fields $\psi_a, \vartheta_a, h_a$ and $w_{am}$ 
we obtain
\begin{equation} \label{S_dual_massless2}
 S_\text{dual} = \ix \big(-\del_{m}f_a\del^{m}f_a^{*}-i\varphi_a\sigma^{m}\del_{m}\bar{\varphi}_a - \cc_a \cc_a\big)\ .
\end{equation}
Just like the original action \eqref{S_kin}, the dual action describes $N_3$ complex scalars and $N_3$ Weyl spinors. The field strengths of the 3-forms are
replaced by the constants $\cc_a$, that also create a constant positive potential. In fact, the superfield equation of motion \eqref{eom_F} includes the duality relation
\begin{equation}
  H^a =  \delta^{a\bb}\, \Im\, n^*_\bb = -\delta^{ab}\cc_b,
\end{equation}
so that the cosmological constants of action and dual action coincide.

Before we proceed let us note that
in the dualization of the massless action a new multiplet $F$ appeared. It differs from the complex linear multiplet described in Appendix \ref{complex_linear} only by the free constant $\cc$ (for $\cc=0$ they coincide). This difference arises 
from the fact that $S$ is not a general chiral superfield
but constructed from a real superfield $U$ via \eqref{defS}.
 If $U$ was complex then $\bar D^2 F_a$ and $D^2 \bar F_a$ had to vanish 
separately in \eqref{eomU_massl} as in the known duality between the chiral and the complex linear multiplet \cite{nonmsca}.

\subsection{Dualization of the massive action} \label{ch_dual_renorm_massive}

Let us now determine the dual action of $N_3$ massive 3-form multiplets.
In this case the first order action is given by
\begin{equation} \label{S_first_renorm_massive}
 S_\text{first} = \ivoll \big(- \delta^{a\bb}F_a \bar F_\bb + F_a S^a + \bar{F}_\ba \bar{S}^\ba - \tfrac{1}{2} m_{ab}^{2} U^{a}U^{b} + \xi_{a}U^{a}\big)\ ,
\end{equation}
where we simply added the mass and Fayet-Iliopoulos terms to \eqref{S_first_renorm}.  Since the equations of motion for $F_a$ are the same as in the massless case, the massive action \eqref{actionmassiv} is correctly reproduced when the $F_a$ are eliminated from \eqref{S_first_renorm_massive}.

In order to find the dual action we first rewrite \eqref{S_first_renorm_massive}
as in \eqref{S_first2}
\begin{equation} \label{S_first_renorm_massive2}
 S_\text{first} = \ivoll \big(- \delta^{a\bb}F_a \bar F_\bb -
 \tfrac{1}{4}U^a\l(\bar{D}^{2} F_a + D^{2}\bar F_a\r) -
 \tfrac{1}{2}m_{ab}^{2}U^{a}U^{b} + \xi_{a}U^{a}\big)\ ,
\end{equation}
and then again 
eliminate the 3-form multiplets $U^a$ by their equations of motion
\begin{equation} \label{eomU_massive}
 -\tfrac{1}{4}(\bar{D}^{2} F_a + D^{2}\bar F_a) - m^2_{ab}U^b + \xi_a = 0\ .
\end{equation}
In contrast to the massless 
case \eqref{eomU_massl} the superfields $F_a$ now remain unconstrained
and the complex scalars $d_a, n_a$ and the Weyl spinors $\zeta_a$ no longer drop out of the dual action.
Substituting \eqref{eomU_massive} into \eqref{S_first_renorm_massive2} and 
using the abbreviation 
\begin{equation}\label{def_Omega}
\Omega_a := \tfrac{1}{4}\l(\bar{D}^{2} F_a + D^{2}\bar F_a\r)\ , 
\end{equation}
we obtain the dual action
\begin{equation}
 S_\text{dual} = \ivoll \big(-\delta^{a\bb}F_a \bar F_\bb + \tfrac{1}{2} m^{-2ab}(\Omega_a - \xi_a)(\Omega_b - \xi_b) \big).
\end{equation}
Using the $\theta$-expansions of $F_a$ and $\Omega_a$ as given in \eqref{defF} and \eqref{eomU_massl} respectively, we find the component action
\begin{equation} \label{S_dual_renorm_mass}
\begin{aligned} S_\text{dual} = \ix \Big(&-f_a d_a^* - f_a^* d_a - \tfrac{1}{2}\del_m f_a \del^m f_a^*
	 + \tfrac{i}{2}\del_m f_a w_a^{m*} - \tfrac{i}{2}w_a^{m}\del_{m}f_a^{*} + \tfrac{1}{2} w_a^{m} w_{am}^{*} \\
	& + \tfrac{1}{2}\psi_a\vartheta_a + \tfrac{1}{2}\bar{\psi}_a\bar{\vartheta}_a
	+\tfrac{1}{\sqrt{2}}\varphi_a\zeta_a + \tfrac{1}{\sqrt{2}}\bar{\varphi}_a\bar{\zeta}_a
	- i\varphi_a\sigma^m\del_m\bar{\varphi}_a - h_a h_a^* - n_a n_a^* \\
	& + m^{-2ab}\l(- \del_{m}n_a\del^{m}n_b^* - \tfrac{i}{2} \zeta_a\sigma^{m}\del_{m}\bar{\zeta}_b +
	  d_a d_b^*   \r) \Big)\ . \eal 
\end{equation}
Note that the FI-parameters $\xi_a$ have dropped out of the action due
to the fact that the highest component of $\Omega_a$ is a total
space-time divergence (remember that they also dropped out of the
original action \eqref{actionmassiv} by a field redefinition). The
action \eqref{S_dual_renorm_mass} still contains the auxiliary fields
$h_a,\ \psi_a,\ \vartheta_a,\ d_a$ and $w_a^m$. Eliminating them by
their equations of motion and performing the field redefinitions
\begin{equation}
  n^\prime _a := \delta_{ab}m^{-1bc} n_c\ , \qquad \zeta^\prime_a := -\tfrac{1}{\sqrt{2}} \delta_{ab}m^{-1bc}\zeta_c\ ,
\end{equation}
yields the on-shell action
\begin{equation} \label{S_dual_renorm_mass2}
 \begin{aligned}
S_\text{dual} = \ix \big(& - \del_m f_a \del^m f_a^{*} - m^2_{ab}f_a f_b^* 
- \del_{m}n^\prime_a\del^{m}n_a^{\prime *} -  m^2_{ab} n^\prime_a n_b^{\prime *} \\
 & - i\varphi_a\sigma^{m}\del_{m}\bar{\varphi}_a -i \zeta^\prime_a\sigma^{m} \del_{m}\bar{\zeta}^\prime_a -  m_{ab}(\varphi_a\zeta^\prime_b + \bar{\varphi}_a \bar{\zeta}^\prime_b) \big)\ .
\eal
\end{equation}
The action \eqref{S_dual_renorm_mass2} is dual to the renormalizable massive action of the 3-form multiplet given in \eqref{S_renorm_mass_onshell}
and describes the dynamics of $2N_3$ massive complex scalars $f_a, n^\prime_a$
and $N_3$ massive Dirac spinors $\varphi_a,\zeta^\prime_a$. The massive 3-forms $C^a_{npq}$ and the real scalars $B^a$  that appear in \eqref{S_renorm_mass_onshell} are represented in the dual action by the complex scalars $n^\prime_a$, so that action and dual action again contain an equal number of on-shell degrees of freedom.

\section{Non-renormalizable action}\label{sec:three}

\subsection{From the superfield Lagrangian to the on-shell action}

In this section we drop the requirement of renormalizability and consider an action with
arbitrary real functions  $K(S, \bar S)$ and $G(U-L')$ given
by\footnote{We choose a minus sign for the G-term in order to have
  canonical kinetic terms for the scalars $B^a$ for a positive definite $G_{ab}$.}
\begin{equation} \label{S_sigma}
 S_3 =  \ivoll \Big(K(S, \bar S) - G(U-L') \Big)\ .
\end{equation}
 $S_3$ is invariant under the combined gauge transformations \eqref{gaugeU},
\eqref{gaugeL} and as before we choose the gauge $L'=0$.  For simplicity we restrict our analysis 
to the bosonic part of the action from now on by setting 
all fermionic components to zero. Using 
 \eqref{def_U} and \eqref{S_components}
we obtain the component form of \eqref{S_sigma}
\begin{equation} \bal\label{S_sigma_kin}
 S_3 = \ix \Big( &K_{a \bb} \big(-\del^{m} M^{a} \del_{m} M^{*\bb} + D^aD^b +H^a H^b \big) - 2(\Im K_{a\bb})H^a D^{b} \\
&-G_{a}\,\big(D^{a}-\tfrac{1}{4}\Box B^{a} \big)
	-\tfrac{1}{2}G_{ab}\,\big( 2M^{a}M^{b*}+\tfrac{1}{3}C^{anpq}C^b_{npq} \big)
\Big)\ , \eal
\end{equation}
where we defined
\begin{equation}\bal\label{K_der}
 K_{a_1\dots a_n \bb_1 \dots \bb_m}(M, M^*)\ &:=\ \frac{\del K}{\del S^{a_1} \dots \del S^{a_n} \del\bar S^{\bb_1} \dots \del\bar S^{\bb_m}} \bigg|_{\theta=\btheta=0}\ ,\\[1ex]
G_{a_1\dots a_n }(B)\ &:=\ \frac{\del G}{\del U^{a_1} \dots \del U^{a_n} } \bigg|_{\theta=\btheta=0}\ .
\eal\end{equation}
We see that the complex scalar fields $M^a$ can be viewed as coordinates
of a K\"ahler manifold with metric $K_{a\bar{b}}$ derived from the K\"ahler potential $K$.
$K_{a\bb}$ is taken to be positive definite and hence
also its symmetric part,   which for a K\"ahler metric coincides with
its real part is positive definite and  
in particular invertible.
The equations of motion for the auxiliary fields $D^a$ imply
\begin{equation}\label{eom_D_sigma}
 D^a = (\Re K)^{-1ab}\l(\tfrac{1}{2}G_b - (\Im K)_{bc}H^c\r),
\end{equation}
where $(\Im K)_{bc}$ denotes the imaginary part of the K\"ahler metric
while $(\Re K)^{-1ab}$ denotes the inverse of its real part.
Inserting \eqref{eom_D_sigma} into \eqref{S_sigma_kin}
 we obtain 
\begin{equation}\bal \label{S_sigma_onshell}
S_3 &= \ix \big[ K_{a\bb}\big( -\del^{m}M^{a}\del_{m}M^{*\bb} + H^{a}H^{\bb} \big) - G_{ab}\big(M^{a}M^{b*}+\tfrac{1}{6}C^{anpq}C^b_{npq}\big) \\
	& \hspace{2 cm} +\tfrac{1}{4}G_{a}\Box B^{a} - \tfrac{1}{4}\big(G_a + 2H^c(\Im K)_{ca}\big)(\Re K)^{-1ab}\big(G_b - 2(\Im K)_{bd}H^d\big)\big] \\
	&= \ix \big[ -K_{a\bb}\,\del^{m}M^{a}\del_{m}M^{*\bb}  - G_{ab}\big(\tfrac{1}{4}\del^{m}B^{a}\del_{m}B^b + M^{a}M^{b*} + \tfrac{1}{6}C^{anpq}C^b_{npq}\big) \\
	& \hspace{2 cm} + g_{ab} H^a H^b + G_a(\Re K)^{-1ab}(\Im
        K)_{bc}H^c - \tfrac{1}{4}G_{a}(\Re K)^{-1ab}G_b\; \big]\ ,
\eal\end{equation}
where in the second step integration by parts was used.
Note that $G_{ab}$ as the sigma model metric of $B^{a}$ coincides with the mass
matrix of $M^{a}$ and $C^{a}$.
In \eqref{S_sigma_onshell}
we also defined the real metric
\begin{equation} \label{def_g}
 g_{ab} := (\Re K)_{ab} + (\Im K)_{ac}(\Re K)^{-1cd}(\Im K)_{db} =
 K_{a\bar{c}}(\Re K)^{-1cd}K_{\bar{d}b}\ ,
\end{equation}
where the second expression for $g_{ab}$ shows explicitly that it is positive definite and that its inverse is given by
\begin{equation} \label{g_inv}
 g^{-1ab} = \Re(K^{-1\ba b})\ .
\end{equation}
 The last term in \eqref{S_sigma_onshell} plays the role of a scalar potential. Depending on the choice of the functions $K$ and $G$ it can lead to non-vanishing vacuum expectation values of the fields $B^a$ and $M^a$ as in the renormalizable action (\ref{S_renorm_massive}). 

The non-renormalizable action for massless 3-form multiplets 
can be obtained from
\eqref{S_sigma}, or \eqref{S_sigma_onshell} respectively, by setting
$G=0$. As we learned in section~\ref{RenormMassless} in this case it is important to add appropriate boundary terms to the action \eqref{S_sigma}.
These terms should be such that they cancel all variational boundary terms 
containing $\delta U^a$ in favor of boundary terms containing $\delta
S^a$, which we assume to vanish.

Let us pause and outline the general prescription for
finding the correct boundary terms for an arbitrary gauge invariant action $S_3$ involving the 3-form multiplets
\begin{equation} \label{3form_action}
 S_3 = \ivoll K(S, \bar S, F, \bar F)\ ,
\end{equation}
where $F$ denotes possible other superfields whose variations are assumed to vanish at the boundary.
Each term in the variation of this action with respect to $U^a$
\begin{equation}
 \delta S_3 = -\tfrac{1}{4} \ivoll \l(\der{K}{S^a} \bar D^2(\delta U^a) + \der{K}{\bar S^\ba} D^2(\delta U^a)\r)
\end{equation}
 contains exactly one $\delta U^a$ with two superspace derivatives acting on it.
Thus one has to apply integration by parts twice for each term in $\delta S_3$
\begin{equation}
 -\tfrac{1}{4} \ivoll \bar D_\dalpha \l(\der{K}{S^a} \bar
 D^\dalpha(\delta U^a) - \l(\bar D^\dalpha \der{K}{
   S^a}\r)\;\delta U^a \r) + \hc\ .
\end{equation}
In order to exchange these for terms depending only on $\delta S^a$,
$\bar D^\dalpha (\delta S^a)$
and their complex conjugates, 
one adds to the action \eqref{3form_action} the boundary terms
\begin{equation}
\mathcal B =  \tfrac{1}{4} \ivoll \bar D_\dalpha \l(\der{K}{S^a} \bar D^\dalpha U^a - \l(\bar D^\dalpha \der{K}{S^a}\r) U^a \r) + \hc\ .
\end{equation}
Indeed, the boundary terms in the variation $\delta(S_3 + \mathcal B)$ are then given by
\begin{equation}
 \tfrac{1}{4} \ivoll \bar D_\dalpha \l(\l(\delta\der{K}{S^a}\r) \bar D^\dalpha U^a - \l(\bar D^\dalpha \delta\der{K}{S^a}\r)U^a \r) + \hc\ 
\end{equation}
which vanish by the variational constraints.

Following this prescription we add to the action \eqref{S_sigma} the boundary term
\begin{equation} \label{B_sigma}
\bal \mathcal B &= -\tfrac{1}{4} \ivoll \bar D_\dal \big(\bar D^\dal K_a(S,\bar S) U^a - K_a(S,\bar S) \bar D^\dal U^a\big) + \hc\\
    &= \Re \ivoll \bar D_\dal \l(K_a(S,\bar S) \bar D^\dal U^a\r) - \tfrac{1}{2} \Re \ivoll \bar D^2\l(K_a(S,\bar S) U^a\r)\ .  \eal
\end{equation}
The first term in the second line of \eqref{B_sigma} contains all the boundary terms involving the 3-forms $C^a_{npq}$ without derivatives, which are
\begin{equation} \label{bt_3form_sigma}
\bal
 \mathcal B_3 &= -\tfrac{1}{3} \ix\, \del_m \Big(\big( (\Re K)_{ab}H^b - (\Im K)_{ab}D^b\big) \veps^{mnpq}C^a_{npq} \Big)\\
&= -\tfrac{1}{3} \ix \del_m \big(g_{ab} H^{a} \veps^{mnpq} C^b_{npq} \big)\ ,
\eal
\end{equation}
where we used 
\eqref{eom_D_sigma} in the second line.
Now we are ready to eliminate the 3-forms from the massless sigma model action
given by
\begin{equation} \label{3_form_term}
 S_3 = \ix \big(\!-K_{a\bb}\,\del^{m}M^{a}\del_{m}M^{*\bb} 
+g_{ab} H^a H^b \big) + \mathcal B_3\ .
\end{equation}
The equations of motion for $H^a$ enforce 
\begin{equation}
 g_{ab}H^b = c_a \ ,\qquad \tn{with} \qquad c_a \in \mathbb{R}\ .
\end{equation}
Inserted back into \eqref{3_form_term} one obtains
\begin{equation}\label{sigma_massl}
 S_3 = \ix \big(\!-K_{a\bb}\,\del^{m}M^{a}\del_{m}M^{*\bb} - \Re (K^{-1\ba b})\, c_a c_b \big)\ .
\end{equation}
We see that the massless 3-forms generate a (positive) potential for
the scalars $M^a$. Since $K_{a\bb}$ is positive definite, a positive cosmological constant is induced and supersymmetry is
 spontaneously broken whenever one $c_a \neq 0$.
As we will see below, the same phenomenon occurs in the dual action.

\subsection{Dual action in the massless case} \label{sigma_dual_massl}

We now want to find a dual action for the massless action \eqref{S_sigma}, i.e.\
with $G = 0$.\footnote{In Appendix~\ref{app:real} we discuss the
  special situation of a K\"ahler potential with an additional shift symmetry.} 
For the first order action we make the ansatz
\begin{equation} \label{S_first_massl}
 S_\text{first} = \ivoll \big( -\hK (F,\bar F) + F_{a}S^{a} + \bar F_{\bb} \bar S^{\bb} \big)\ ,
\end{equation}
where $\hK $ is real. The equations of motion for $F_a$ then read
\begin{equation} \label{eomF_sigma}
 \der{\hat K}{F_a} = S^a\ .
\end{equation}
In order to reproduce (\ref{S_sigma}) (with $G = 0$)
$\hat K$ has to fulfill the equation
\begin{equation}
 K\Big(\der{\hK}{F},\der{\hK}{\bar F}\Big) = 
F_a\, \der{\hK }{F_a} + \bar F_\bb\, \der{\hK }{\bar F_\bb} - \hK (F,\bar F)\ .
\end{equation}
This relation is satisfied when $K$ is the Legendre transform of $\hK$ and vice versa (the Legendre transformation
is its own inverse, see App.~\ref{App_Legendre}).
Therefore we assume here that $K$ is strictly convex so that a Legendre transform as defined in Appendix~\ref{App_Legendre} exists. Then the relation \eqref{eomF_sigma} is invertible and equivalent to
\begin{equation}
 F_a = \der{K}{S^a} \equiv K_a(S,\bar S)\ .
\end{equation}

The dual action is obtained by eliminating $U^a$ from the first order
action \eqref{S_first_massl} together with 
the boundary terms \eqref{B_first}. Note that the latter
exactly reproduce the terms given in \eqref{B_sigma} for $F_a = K_a(S,\bar S)$.
Just like in \eqref{S_first2}, we can then write the action in the form
\begin{equation}\label{Sint}
 S_\text{first} + \mathcal B_\tn{first} = \ivoll \Big(-\hK (F,\bar F)- \tfrac{1}{4}U^a(\bar{D}^{2}F_a + D^{2}\bar F_a) \Big)\ .
\end{equation}
As before, variation with respect to $U^a$ yields the constraint \eqref{eomU_massl}
and thus $F_a$ again takes the form (\ref{F_constr}).
Inserted into \eqref{Sint} one obtains the dual action
\begin{equation}\begin{aligned} \label{S_dual_sigma1}
S_{\text{dual}} &= -\ivoll \hK (F, \bar F) \\
	&= - \ix \Big[ \hK^a\big(-\tfrac{1}{4}\Box f_a - \tfrac{i}{2}\del_m w_a^m\big) + \hK^\ba \big(-\tfrac{1}{4}\Box f_\ba^* + \tfrac{i}{2}\del_m w_{\ba}^{m*}\big)\\
      &\qquad\qquad  + \hK^{ab}\big(-\tfrac{1}{4}w_{a m}w_{b}^{m} + i\cc_{a}h_{b}\big) + \hK^{\ba\bb} \big(-\tfrac{1}{4}w_{\ba m}^{*}w_{\bb}^{m*}-i\cc_{\ba}h_{\bb}^{*} \big) \\
     &\qquad\qquad  + \hK^{a\bb}\big(-\tfrac{1}{2}w_{am}w_\bb^{m*} + h_{a}h_\bb^* + \cc_a \cc_\bb \big) \Big]\ .
\end{aligned}\end{equation}
The fields $h_a$ and $w_{am}$ have purely algebraic equations of motion and can thus be eliminated. This is most conveniently done for the $h_a$
by completing the square as described in 
Appendix \ref{elim_aux} with the result\footnote{As $K$ is the Legendre transform of $\hK$, it is implicit in formula \eqref{kaehler_metric} (with $K$ and $\hK$ exchanged) that  $\hK^{a\bb}$ is invertible.} 
\begin{equation}\begin{aligned} \label{S_dual_sigma2}
S_{\text{dual}} = \ix \Big[& \tfrac{1}{4}\hK^{ab}w_{a}^mw_{bm} + \tfrac{1}{4}\hK^{\ba\bb} w_\ba^{m*}w_{\bb m}^* + \tfrac{1}{2}\hK^{a\bb}w_a^m w_{\bb m}^* \\
	& -\del_m \hK^a \l(\tfrac{1}{4}\del^m f_a + \tfrac{i}{2}w_a^m\r) - \del_m \hK^\ba \l(\tfrac{1}{4}\del^m f_\ba^* - \tfrac{i}{2}w_\ba^{m*}\r) \\
	 & +\big(\hK^{ac}\hK^{-1}_{c\bar d}\hK^{\bd\bb}-\hK^{a\bb}\big)\cc_a\cc_\bb 
\Big]\ .
\end{aligned}\end{equation}
Following the prescription given in Appendix \ref{elim_aux}, we can also complete the square with respect to the fields $w_a^m$ by writing the action in the form
\begin{equation} \label{S_dual_square}
 \bal S_{\text{dual}} = \ix &\Big[ \tfrac{1}{4}\big((w_a^m + u_a^m)\ (w_{\ba}^{m*} + u_{\ba}^{m*})\big) \begin{pmatrix} \hK^{ab} & \hK^{a\bb} \\ \hK^{\ba b} & \hK^{\ba\bb} \end{pmatrix} \begin{pmatrix} w_{bm} + u_{bm} \\ w^*_{\bb m} + u^*_{\bb m} \end{pmatrix} \\
 & -\tfrac{1}{4}\big(u_a^m\ u_{\ba}^{m*})\begin{pmatrix} \hK^{ab} & \hK^{a\bb} \\ \hK^{\ba b} & \hK^{\ba\bb} \end{pmatrix} \begin{pmatrix} u_{bm} \\ u^*_{\bb m} \end{pmatrix} - \tfrac{1}{4}\del^m\! f_a\, \del_m \hK^a - \tfrac{1}{4} \del^m\!f_{\ba}^*\,\del_m \hK^{\ba} \\
 &  +\big(\hK^{ac}\hK^{-1}_{c\bar d}\hK^{\bd\bb}-\hK^{a\bb}\big)\cc_a\cc_\bb \Big]\ ,  \eal
\end{equation}
where the $u_a$ solve the equations
\begin{equation}
\begin{pmatrix} \hK^{ab} & \hK^{a\bb} \\ \hK^{\ba b} & \hK^{\ba\bb} \end{pmatrix} \begin{pmatrix} u_{bm} \\ u^*_{\bb m} \end{pmatrix} = -i \begin{pmatrix} \del_m\hK^a \\ -\del_m\hK^{\ba} \end{pmatrix}\ .
\end{equation}
The action \eqref{S_dual_square} depends on 
the Hesse matrix of the Legendre transformed K\"ahler potential $\hK(f,f^*)$
which, as derived in Appendix \ref{App_Legendre}, is the inverse of the Hesse matrix of $K(M, M^*)$. However, let us ignore this fact for the moment and only notice that $\Hess \hK$ is invertible
with its inverse  given by (cf.\ \eqref{H_inv})
\begin{equation} \label{Hess_inv}
 (\Hess \hK)^{-1} = \begin{pmatrix} C & D \\ D^* & C^* \end{pmatrix} \qquad \tn{where} \qquad 
  \bal D_{a\bb} &= \l(\hK^{\bb a} -  \hK^{\bb\bar{c}}\hK^{-1}_{\bar{c}d}\hK^{da}\r)^{-1}\ ,\\
       C_{ab} &= - \hK^{-1}_{a\bar{c}}\hK^{\bar{c}\bar{d}}D^*_{\bar{d}b}\ . \eal
\end{equation}
 The equations of motion for the $w_{am}$ and $w^*_{\ba m}$ imply that the term in the first line of \eqref{S_dual_square} (the ``square'') vanishes
and thus we obtain the on-shell action
\begin{equation} \label{S_dual_sigma_onshell_1}
 \bal S_{\text{dual}} = \ix \Big[ & -\tfrac{1}{2} \big(\del^m\hK^a D_{a\bb} \del_m\hK^{\bb}\big) + \tfrac{1}{4}\big(\del^m\hK^a C_{ab}\del_m\hK^b +  \del^m\hK^{\ba} C^*_{\ba\bb}\del_m\hK^{\bb} \big) \\
 &  - \tfrac{1}{4}\del^m f_a \del_m \hK^a - \tfrac{1}{4} \del^m f_{\ba}^* \del_m \hK^{\ba}  + \l(\hK^{ac}\hK^{-1}_{c\bar d}\hK^{\bar d \bar b}-\hK^{a\bb}\r)\cc_a\cc_\bb 
\Big]\ . \eal
\end{equation}
To simplify this expression, we substitute
$ C_{ab} = - D_{b\bc}\hK^{\bc\bar{d}}\hK^{-1}_{\bar{d}a}$,
use the chain rule for $\del^m\hK^a$ and the relation
\begin{equation}
 \hK^{\bc\bd}\hK^{-1}_{\bd a}\hK^{ae} = \hK^{\bc e} - D^{-1\bc e}\ .
\end{equation}
This leads to
\begin{equation}\label{S_dual_sigma_onshell_2}
 \bal S_{\tn{dual}} = \ix \big(\! -D_{a\bb}\, \del^m\hK^a \del_m\hK^{\bb} - \Re(D^{-1\bb a})\cc_a\cc_b \big)\ . \eal
\end{equation}
Using $(\Hess \hK)^{-1} = \Hess K$ we obtain from \eqref{Hess_inv} (cf.\ \eqref{kaehler_metric})
\begin{equation} \label{rel_D-K}
 D_{a\bb}(f,f^*) = K_{a\bb}(M,M^*)\ ,
\end{equation}
where $K_{a\bar b}$ has to be evaluated at 
\begin{equation} \label{Legendre_lc}
 M^a  = \hK^a(f,f^*)\ , \qquad M^{*\ba} = \hK^{\ba}(f,f^*)\ .
\end{equation}
Equation \eqref{Legendre_lc} is just the lowest component of the
Legendre relation \eqref{eomF_sigma} that appeared as the equation of
motion for $F_a$ in the first order action
\eqref{S_first_massl}. Using \eqref{rel_D-K} and 
\eqref{Legendre_lc}
we see that the kinetic term of the dual on-shell action
\eqref{S_dual_sigma_onshell_2} is equal to that of the original action
\eqref{S_sigma_kin}. In particular the ``new'' metric
\begin{equation}\label{newmetric}
 D_{a\bb} = \l(\hK^{\bb a} -  \hK^{\bb\bar{c}}\hK^{-1}_{\bar{c}d}\hK^{da}\r)^{-1}
\end{equation}
appearing in the dual action is again K\"ahler with respect to its natural variables $\hK^a$ and $\hK^\bb$.\footnote{For the case of the complex linear multiplet, i.e.\ for $\cc_a=0$, ref.\ \cite{comments} derives a different result 
that apparently does not lead to a K\"ahler geometry in the dual action.
We thank J.\ Gates for discussing this issue.}

Let us now show
that also the potential coincides upon using \eqref{Legendre_lc} and the duality relation of the constants $c_a$ and $\cc_a$ which is contained in \eqref{eomF_sigma}.  The $\thz$ and $\thq$-components of  \eqref{eomF_sigma} with constrained $F_a$ (i.e., $n_a = i\cc_a$) read
\begin{equation} \bal \label{eomF_theta_theta}
 \hK^{ab}h_b - i\hK^{a\bb}\cc_\bb = D^a + iH^a\ ,\qquad
 \hK^{a\bb}h_\bb^* + i\hK^{ab}\cc_b = 0\ .
\eal \end{equation}
The second equation in \eqref{eomF_theta_theta} is just the equation of motion for the auxiliary field $h_a$ which we already used to compute the on-shell action \eqref{S_dual_sigma_onshell_2}. Inserting the solution for $h^*_\bb$
into the complex conjugate of the first equation in \eqref{eomF_theta_theta} and using \eqref{newmetric}, one finds the on-shell duality relation
\begin{equation} \label{dual_H-c}
\bal H^a &= \Re \big(\hK^{ab}\hK^{-1}_{b\bar{c}}\hK^{\bar{c}\bar{d}} - \hK^{a\bar{d}} \big)\cc_d 
  = -\Re\big(D^{-1a\bar{d}}\big)\,\cc_d\ . \eal
\end{equation}
From this last equation, \eqref{g_inv} and \eqref{rel_D-K} one can derive the relation between the constants $c_a$ and $\cc_a$ appearing in the on-shell action \eqref{sigma_massl} 
and dual action \eqref{S_dual_sigma_onshell_2} to be
\begin{equation} \label{dual_cc}
 c_a = g_{ab}H^b =  -\cc_a\ .
\end{equation}
Now we see that the sigma model action of the massless 3-form multiplet  \eqref{sigma_massl}, is indeed equal to its dual action \eqref{S_dual_sigma_onshell_2} by use of the two duality relations \eqref{Legendre_lc} and \eqref{dual_cc}.

In conclusion, we discuss the relation of the action \eqref{sigma_massl} with its dual \eqref{S_dual_square}.
The two superfield equations of motion of the first order action
\eqref{eomF_sigma} and \eqref{eomU_massl} can be used to eliminate the
corresponding superfields  from the action so that it becomes the
original action $S_3$ of the 3-form multiplet or the dual action
$S_\tn{dual}$ respectively. Together they contain all  equations of
motion of the first order action, in particular also those of the
auxiliary fields which where used to eliminate them from action and
dual action to obtain the final on-shell actions \eqref{sigma_massl}
and \eqref{S_dual_sigma_onshell_2}. Thus it is clear that one can
translate these on-shell actions into each other by making use of all
the information contained in \eqref{eomF_sigma} and
\eqref{eomU_massl}. 
However, the situation is not that simple because these two superfield
equations also contain the equations of motion of the physical fields
$M^a$ and $f_a$ appearing in the on-shell action and dual
action. Therefore one might not expect that these actions can be
translated into each other only by use of the duality relations
\eqref{Legendre_lc} and \eqref{dual_cc}. It has to be considered as
coincidence that this is nevertheless possible for the massless case,
so that the transition from action to dual action can be described as
a simple field redefinition. Clearly, it would not be possible if the
number of off-shell degrees of freedom of action and dual action did
not coincide. (In the massive  case they do not coincide and one has
to make use of the equations of motion of the physical fields to
translate the dual action back into the 3-form action.) On the other
hand, using the duality relations \eqref{Legendre_lc} and
\eqref{dual_cc} as a field redefinition to re-express the dual action
in terms of the fields $M^a$ and constants $c_a$ one will find an
action whose equations of motion are equivalent to those of the dual
action by these relations. As the massless 3-form action
\eqref{sigma_massl} has the same property, it is not very surprising
that they coincide. 

\subsection{Dual action in the massive case} \label{sigma_dual_massive}

Let us now turn to the massive case where the potential $G(U)$ is nontrivial.
We simply add this term to the first order action \eqref{S_first_massl}
and consider
\begin{equation} \label{S_first_massive}
 S_\text{first} = \ivoll \Big( -\hK (F, \bar F) + F_{a}S^{a} + \bar F_\bb \bar S^\bb - G(U) \Big)\ ,
\end{equation}
where $\hK$ is again the Legendre transform of $K$. Since the Euler-Lagrange equations \eqref{eomF_sigma} do not change, the original action \eqref{S_sigma} is reproduced correctly. We then rewrite the action as\footnote{Again, in the massive case it is legitimate to drop boundary terms.}
\begin{equation}
 S_\text{first} = \ivoll \Big(-\hK (F, \bar F) - \tfrac{1}{4}U^a (\bar{D}^{2}F_a + D^{2}\bar F_a) - G(U) \Big)\ ,
\end{equation}
and determine the Euler-Lagrange equation for $U^a$ to be 
\begin{equation} \label{eomU_sigma_massive}
	 \der{G}{U^a}  \; = \; -\tfrac{1}{4}(\bar{D}^{2}F_a + D^{2} \bar F_a) = \Omega_a\ .
\end{equation}
To eliminate the $U^a$ from the action we have to assume that $G$ has a Legendre transform $\hat G$. When this is the case, we find as a dual action
\begin{equation}
 S_\tn{dual} = \ivoll \big(-\hK (F,\bar F) + \hat G(\Omega)\big)\ .
\end{equation}
Using \eqref{defF} and \eqref{eomU_massl} we obtain the component action
\begin{equation}\label{Sd}
 \bal S_\tn{dual} 
    &= - \ix \Big[ \hK^a \l(d_a - \tfrac{1}{4}\Box f_a - \tfrac{i}{2}\del_m w_a^m\r) + \hK^{ab}\l(- \tfrac{1}{4}w_a^m w_{b m} + h_a n_b \r) + \tn{h.c.}\\
    &\hspace{2cm} + \hK^{a \bb}\l(- \tfrac{1}{2} w_a^m w_{\bb m}^* + h_a h_\bb^* + n_a n_\bb^* \r) + \hG^{ab}\l(d_a d_b^* - \del_m n_a \del^m n_b^*\r) \Big]\ . \eal
\end{equation}
Note that compared to \eqref{S_dual_sigma1}, the complex scalars $d_a$
and $n_a$ also appear in the massive dual action since now the $F_a$ remain unconstrained.
The equations of motion for the auxiliary fields $d_a$ and $h_a$ implied by 
\eqref{Sd} are 
\begin{equation}
  -\hK^a + \hG^{ab}d_b^* = 0\ ,  \qquad
    \hK^{ab}n_b + \hK^{a\bb}h_{\bb}^* = 0\ .  
\end{equation}
Inserted into \eqref{Sd}
yields\footnote{Note that $\hG^{ab}$ is invertible with $\hG^{-1}_{ab} = G_{ab}$}
\begin{equation} \label{S_dual_sigma_mass}
 \begin{aligned} S_\tn{dual} = \ix \Big[& \tfrac{1}{4}\hK^{ab}w_{a}^mw_{bm} + \tfrac{1}{4}\hK^{\ba\bb} w_\ba^{ m*}w_{\bb m}^* + \tfrac{1}{2}\hK^{a \bb}w_a^m w_{\bb m}^* \\
	& -\del_m \hK^a \l(\tfrac{1}{4}\del^m f_a + \tfrac{i}{2}w_a^m\r) - \del_m \hK^\ba \l(\tfrac{1}{4}\del^m f_{\ba}^* - \tfrac{i}{2}w_{\ba}^{m*}\r)  \\
& - D^{-1a\bb} n_a n_\bb^* -  \hK^{\bb}\hG^{-1}_{ba}\hK^a - \hG^{ab} \del^m n_a \del_m n_b^* \Big]\ , \eal
\end{equation}
where $D^{-1a\bb} = \hK^{a\bar b} - \hK^{ac}\hK^{-1}_{c \bar d}\hK^{\bar d \bb}$.
Note that the action \eqref{S_dual_sigma_mass} differs from the massless action
\eqref{S_dual_sigma2} only by the three terms given in  the last line 
of \eqref{S_dual_sigma_mass}. Therefore 
the auxiliary fields $w_a^m$ can be eliminated using 
the same steps as in the massless case and we obtain 
the massive dual action
\begin{equation} \label{S_dual_sigma_mass_final}
S_\tn{dual} = \ix \Big( -D_{a\bb}\del^m \hK^a \del_m \hK^{\bb} -  \hG^{ab} \del^m n_a \del_m n_b^*
 - D^{-1a\bb}n_a n_\bb^* - \hK^a \hat G^{-1}_{ab}\hK^\bb \Big)\ .
\end{equation}
Just as for the renormalizable action discussed in section
\ref{ch_dual_renorm_massive}, the massive 3-forms are no longer dual
to  constants but are 
represented, together with the real scalars $B_a$, by the complex
scalars $n_a$.
Note that $\hat G^{ab}$ 
can be viewed as a  K\"ahler metric derived from the K\"ahler potential
$\hat G(n,n^*)=\hat G(n+n^*)$ which has a shift symmetry.
  The last two terms in
\eqref{S_dual_sigma_mass_final} form the potential of the scalars $f_a$ and $n_a$.

\section{Coupling to chiral fields} \label{ch_coupling}

\subsection{Renormalizable couplings} \label{ch_renorm_coupling}

In this section we study the coupling of $N_3$ 3-form multiplets $U^a$
to $N_c$ chiral multiplets $\Phi^i$. We start with the renormalizable
massive action \eqref{actionmassiv} and add kinetic and interaction
terms  for $\Phi^i$. The action is then of the form\footnote{Note that
  this is a gauge fixed action, with the gauge specified in
  section~\ref{RenormMassive}. Furthermore,  for renormalizable theories 
$m^2_{ab}$ has to be constant and we do not consider a $\Phi U$ coupling because it can be rewritten as a $\Phi S$ coupling in the superpotential using $\idtq \Phi U = -\frac14 \bar D^2(\Phi U) + \tn{tot. divergence}$.}
\begin{equation}
\bal
  S = \ix \Big[\idtdtq \big( S^{a}\bar S^{a} + \Phi^i\bar\Phi^i -
  \tfrac{1}{2} m^{2}_{ab} U^{a} U^b + \xi_a U^a \big)  + \big(\idt
  W(S,\Phi)+ \hc \big)
\Big]\ ,
\eal
\end{equation}
where we split $W$ as
\begin{equation} \label{form_W}
 W(S, \Phi) = W^\tn{int.}(S, \Phi) + W^S(S) + W^{\Phi}(\Phi)\ .
\end{equation}
In order for the action to be renormalizable $W$ can be at most cubic in the superfields.
$\Phi$ has  components $A, \psi$ and $F$ defined
in Appendix \ref{chiral_multiplet} while for  $S$ we use
\eqref{S_components}.
This yields (cf.\ \eqref{actionmassiv})
\begin{equation} \label{S_coupl_renorm}
\bal
 S = \ix \Big[ &-\del^{m}M^{a}\,\del_{m}M^{a*}-i\lambda^{a}\sigma^{m}\del_{m}\bar{\lambda}^{a}  -\del_{m}A^i\,\del^{m}\!A^{i*} - i\psi^i\sigma^{m}\del_{m}\bar{\psi}^i \\
	&+ D^{a}D^{a} + H^{a}H^{a} + F^{i}F^{i*} - m^{2}_{ab}\big(\tfrac{i}{2}\chi^{a}\sigma^{m}\del_{m}\bar{\chi}^b
	-\tfrac{i}{2}\chi^{a}\lambda^b - \tfrac{i}{2}\bar{\chi}^a\bar{\lambda}^b + M^{a}M^{b*} \\
	& \hspace{3cm} + B^aD^b - \tfrac{1}{4}B^a\Box B^b + \tfrac{1}{6}C^a_{npq}C^{bnpq} \big)
	+ \xi_a \big(D^{a}-\tfrac{1}{4}\Box B^{a}\big) \\
	&+ \big( W_a\l(D^a+iH^a\r) + W_i F^i -\tfrac{1}{2}W_{ij}
        \psi^i\psi^j - \tfrac{1}{2}W_{ab}\lambda^a\lambda^b -
        W_{ai}\lambda^a\psi^i + \tn{h.c.} \big) \Big]\ .
\eal
\end{equation}
Compared with the action \eqref{S_superpot_mass}, here we have the additional auxiliary fields $F^i$ that create a contribution $W_i W_i^*$ to the scalar potential.
Thus, after eliminating the auxiliary fields, redefining $B$ and $\chi$ as in \eqref{B_chi_redef} and dualizing the 3-forms to scalars
\begin{equation}
 \phi^a := m^{-1ab}\big(H_b - \Im (W_b)\big)\ ,
\end{equation}
we obtain the action
\begin{equation}\label{S_coupl_renorm_onshell}
\bal
 S = \ix \big[\!&-\del_{m}A^i\,\del^{m}\!A^{i*} - i \psi^{i}\sigma^{m}\del_{m} \bar{\psi}^{i} -\del^{m}M^{a}\,\del_{m}M^{a*}-i\lambda^{a}\sigma^{m}\del_{m}\bar{\lambda}^{a} - i\chi^{\prime a} \sigma^{m}\del_{m}\bar{\chi}^{\prime a}\\[-2mm]
  &-\! \big(m_{ab}\chi^{\prime a}\lambda^b  + W_{ai}\lambda^a\psi^i  + \tfrac{1}{2}W_{ab}\lambda^a\lambda^b + \tfrac{1}{2}W_{ij} \psi^i\psi^j  + \hc \big) - \del_m N^a \del^m\!N^{a*} \! -\! \mathcal V \big]\ .
\eal
\end{equation}
As in section \ref{sec:superpot} we defined $N^a := -B^a + i\phi^a$ and the scalar potential is given by
\begin{equation} \label{potential}
\mathcal V \ = \ m^{2}_{ab} M^{a}M^{b*} + W_i W^*_i + \big(m_{ab}N^b + W_a\big)\big(m_{ac}N^{c*} + W_a^*\big)\ .
\end{equation}
The form of this potential implies that supersymmetry is unbroken if and only if there is a field configuration for which the equations
\begin{equation} \label{susy_cond}
\bal m_{ab}N^b + W_a =0\ ,\qquad
 W_i =0\ ,\qquad M^a =0\eal
\end{equation}
are fulfilled. 
If $m_{ab}$ is invertible
the first equation in \eqref{susy_cond} always has a solution 
which determines the $N^b$.
For renormalizable interactions we can assume that $W^\tn{int.}$
in \eqref{form_W} is of the form
\begin{equation}
W^\tn{int.}(S, \Phi) =  \mu_{a,i} S^a\Phi^i +
 \rho_{a,ij}S^a\Phi^i\Phi^j + \gamma_{ab,i}S^aS^b\Phi^i\ ,
\end{equation}
and as consequence the second and third equation in \eqref{susy_cond}
can be summarized as
\begin{equation}\label{susycond}
 W^\Phi_i(A)= W_i\vert_{M=0} = 0 \ .
\end{equation}
This implies that in the class of models considered here supersymmetry
is broken for exactly the same O'Raifeartaigh superpotentials $W^\Phi$
as in the well known chiral theories \cite{O'Raifeartaigh:1975pr}. 
It is particularly interesting that supersymmetry cannot be broken by 
$W^\tn{int.}$ and $W^S$ for a non-singular mass matrix $m_{ab}$
as the first equation in \eqref{susy_cond} always has a solution.
This also prevented the Fayet-Iliopoulos term from breaking supersymmetry in \eqref{actionmassiv}.

Let us now analyze the mass spectrum of the theory for the special case 
$W^S(S) = 0 = W^\Phi(\Phi)$. 
Note that in this case supersymmetry is unbroken since the potential \eqref{potential} vanishes for $M^a = N^a = A^i = 0$.  
Since all fields have vanishing vacuum expectation values, contributions to the mass matrices only come from terms that are quadratic in the fields.\footnote{We could also include terms that are linear in $S$ as they would only shift the VEVs of the $N^a$. By redefining $N^a \to \langle N^a \rangle + N^a$, we would reduce the problem to the case where $W^S(S) = 0$.}
For the scalars  $M^a$ these can be found in
\begin{equation}\label{mass_M}
 m^{2}_{ab} M^{a*}M^{b} + W^*_iW_i  =  \l(m^2_{ab} + \mu^*_{a,i}\mu_{b,i}\r)M^{a*}M^{b} + \ldots,
\end{equation}
where dots denote terms that are at least cubic.
Next we collect the mass terms for the scalars $N^a$ and $A^i$ which are entirely contained in the term
\begin{equation}
 \l| m_{ab}N^b + W_a \r|^2 = \l|m_{ab}N^b + \mu_{a,i}A^i\r|^2 + \ldots = \big(N^{b*} \ A^{j*} \big) \begin{pmatrix} m_{ba}  \\ \mu_{a,j}^*  \end{pmatrix} \big(m_{ab} \ \mu_{a,i} \big) \begin{pmatrix} N^b \\ A^i \end{pmatrix} + \ldots \ .
\end{equation}
Thus we have determined the quadratic mass matrix for the $A^i$ and $N^a$ which can be written in  matrix notation as
\begin{equation}
 m^2_{N,A} = \begin{pmatrix} m \\ \mu^\dagger \end{pmatrix} \big(m \ \mu \big) = \begin{pmatrix} m^2 & m\mu\\ \mu^\dagger m& \mu^\dagger \mu  \end{pmatrix} \ .
\end{equation}
Obviously, $m^2_{N,A}$ has (at least) $N_c$ zero eigenvalues corresponding to the $N_c$ linearly independent vectors in the kernel of the $N_3 \times (N_3+N_c)$ matrix $(m \ \mu)$. The remaining $N_3$ eigenvalues of $m^2_{N,A}$ coincide with the eigenvalues of the hermitian $N_3 \times N_3$ matrix
\begin{equation}
 Q := \big(m \ \mu \big) \begin{pmatrix} m \\ \mu^\dagger \end{pmatrix} = m^2 + \mu\mu^\dagger \ .
\end{equation}
The corresponding eigenvectors of $m^2_{N,A}$ are given by
\begin{equation}
 \begin{pmatrix} m \\ \mu^\dagger \end{pmatrix} v_a = \begin{pmatrix} m v_a  \\ \mu^\dagger v_a\end{pmatrix}
\end{equation}
when $v_a$, $a=1,\dots,N_3$, are the eigenvectors of  $Q$. Note that the quadratic mass matrix for the $M^a$ given in \eqref{mass_M} coincides with $Q^*$ so that for each eigenvalue of $Q$ there are two complex scalar fields (i.e., four on-shell degrees of freedom) with this mass.\footnote{This is special for the case of vanishing superpotentials. It is generally not true when $W^S$ or $W^\Phi$ is nontrivial, e.g.~for $W^\Phi = \frac12 \tilde\mu_{ij}^2 \Phi^i\Phi^j$.}
Let us also analyze the fermionic mass spectrum in order to determine the physical supermultiplets into which the mass eigenstates are organized.
The quadratic mass matrix for the fermions of the theory, $\lambda^a,
\chi^a$ and $\psi^i$, is found to be
\begin{equation}
 m^2_{\lambda,\chi,\psi} = \begin{pmatrix} 0 & m & \mu^*\\ m &0&0\\ \mu^\dagger & 0 & 0 \end{pmatrix} \begin{pmatrix} 0 & m & \mu\\ m &0&0\\ \mu^T & 0 & 0 \end{pmatrix} = \begin{pmatrix} m^2 + \mu^*\mu^T & 0 & 0 \\ 0 & m^2 & m\mu \\ 0 & \mu^\dagger m & \mu^\dagger \mu \end{pmatrix}\ .
\end{equation}
The $\lambda^a$ mix among themselves to form mass eigenstates and
their mass matrix coincides with that of the $M^a$. As the
corresponding linear combinations of the $\lambda^a$ and $M^a$ are
contained in the same chiral superfield (which is a linear combination
of the $S^a$), it is clear that also the quanta associated to these
fields reside in one chiral supermultiplet. Furthermore, the mass
matrix for the $\chi^a$ and $\psi^i$ is identical with
$m^2_{N,A}$. Thus there are also $2N_c$ massless fermionic states,
which join the $2N_c$ massless bosonic states to form $N_c$ massless
chiral multiplets. Furthermore, there are 
another $2N_3$ fermionic mass eigenstates for the
eigenvalues of $Q$ which reside in $N_3$ chiral multiplets with the
corresponding linear combinations of the $N^a$ and $A^i$ (or, more
precisely, the physical states associated to these fields). 


\subsection{Non-renormalizable coupling} \label{non-ren_coupl}

In the non-renormalizable case we allow for arbitrary couplings
between the $U^a$ and $\Phi^i$  as well as non-renormalizable kinetic
terms that mix  $\Phi^i$ 
and $S^a$. Thus we start with the expression
\begin{equation}
 \bal S = \ix \bigg[ &\idtdtq \Big(K(S, \bar S,\Phi,\bar\Phi) - G(U, \Phi, \bar\Phi) \Big)
+\idt W(S,\Phi)+\idtq  W^*(\bar S, \bar\Phi) \bigg]\ .  \eal
\end{equation}
We again consider only the bosonic part of the action, which has the component form\footnote{Here integration by parts was applied to rewrite the terms with d'Alambert operators. Note that the `mixed kinetic' terms proportional to $\del_m A^i \del^m B^a$ (and complex conjugate), that arise in this step, cancel due to the different signs of the $\Box$-terms in $U$ and $\Phi$. One can argue that such terms cannot appear in a supersymmetric theory as their supersymmetry variation cannot be canceled by the variation of a kinetic term for their superpartners $\psi$ and $\chi$ because $B$ is real while $A$ is complex.}
\begin{equation} \label{S_coupl_comp}
 \bal S = \ix \Big[ &K_{a\bb}\big(\!\!-\del_{m}M_{a}\,\del^{m}\!M^{*\bb}\! +\! D^a D^\bb\! +\! H^a H^\bb \big) + 2(\Im K_{a\bb})H^a D^b \\
  &+ P_{i\bj}\big(\!\! -\!\del_{m}A^i\del^m\!A^{*\bj}\! +\! F^i F^{*\bj} \big) + \big[K_{\ba i}\big(\!\!-\! \del_m M^{*\ba}\del^m\!A^i \!+\!(D^\ba\!-\! iH^\ba)F^i \big)\! +\! \tn{h.c.}\big] \\
  &  
- G_{ab}\big(\tfrac{1}{4}\del_m B^a \del^m\!B^b + M^a\!M^{b*} + \tfrac{1}{6}C^a_{npq}C^{bnpq}\big) \\
&  - G_a D^a - G_{ai}M^a F^i - G_{a\bj}M^{a*}F^{*\bj} 
 + \tfrac{1}{6}i\big(G_{ai}\del^m\!A^i - G_{a\bj}\del^m\!A^{*\bj}\big) \veps_{mnpq}C^{anpq}\\
&  + W_i F^i  + W^*_\bi F^{*\bi} + W_{a}\big(D^{a} + iH^{a}\big) + W^*_\ba
\big(D^\ba -iH^\ba \big) \Big]\ , \eal
\end{equation}
where we defined  $P_{i\bj} := K_{i\bj} - G_{i\bj}$.
The action \eqref{S_coupl_comp} still contains  the auxiliary fields
$F^i$ and $D^a$. To eliminate them, we once more follow the prescription given in Appendix \ref{elim_aux} and first consider only the part of the Lagrangian containing the $F^i$. With the definitions
\begin{equation} \bal
 Z_i := -iH^\ba K_{\ba i} - G_{ai}M^a  + W_i\ , \qquad
 J_i := D^{\ba} K_{\ba i} + Z_i\ ,
\eal \end{equation}
this is\footnote{Since  $P_{i\bj}$ is the sigma model metric for the
  scalars $A^i$ we assume it to be invertible.}
\begin{equation} \label{Lag_F}
 \bal \Lag_{F} &= F^i P_{i\bj}F^{*\bj} + J_i F^i + J^*_\bj F^{*\bj}\\
    &= \big(F^i + J^*_{\bar k}P^{-1
      \bar{k}i}\big)P_{i\bj}\big(F^{*\bj} + P^{-1 \bj k}J_k\big) -
    J^*_\bj P^{-1\bj i} J_i\ . \eal
\end{equation}
After elimination of the $F^i$ only the second term in the second line of \eqref{Lag_F} survives. Then the part of the Lagrangian containing the fields $D^a$ becomes (note that $J_i$ also depends on the $D^a$)
\begin{equation} \label{Lag_D}
 \begin{aligned} \Lag_D &= D^a(K_{a\bb}-K_{a\bj}P^{-1\bj i}K_{i\bb})D^{\bb} + Q_a D^a\\
   &= (D^a + \tfrac{1}{2}Q_{c}R^{-1ca})R_{ab}(D^b + \tfrac{1}{2}R^{-1bd}Q_d) - \tfrac{1}{4}Q_a R^{-1ab}Q_b
 \end{aligned}
\end{equation}
where
\begin{equation} \begin{aligned}
 Q_a &:=  -2(\Im K_{a\bb})H^b - G_a + 2\Re\big(W_a - K_{a\bj}P^{-1\bj i}Z_i\big)\ ,\\
 R_{ab} &:= \Re\big(K_{a\bb}-K_{a\bj}P^{-1\bj i}K_{i\bb}\big)\ .
\end{aligned}\end{equation}
Now the $D^a$ can also be eliminated from the action: again the first term in the second line of \eqref{Lag_D} vanishes and one finds that all the terms in the action \eqref{S_coupl_comp} containing the auxiliary fields $F^i$ and $D^a$ are replaced by
\begin{equation}
 \ix \big(\!-Z^*_{\bj}P^{-1\bj i}Z_i - \tfrac{1}{4}Q_{a} R^{-1ab}Q_b \big)\ .
\end{equation}
This expression contains potential terms for the scalars $M^a, B^a$ and $A^i$ as well as terms involving the field strengths $H^a$. To separate them, let us define
\begin{equation}
 \bal \tilde Z_i &:= - G_{ai} M^a  + W_i\ , \\
    \tilde Q_a &:= - G_a + 2\Re\big(W_a - K_{a\bj}P^{-1\bj i}\tilde Z_i\big)\ . \eal
\end{equation}
Then the on-shell action can be written as
\begin{equation} \label{S_coupl_onshell}
 \bal S = \ix \bigg[ &-K_{a\bb}\del_{m}M^{a}\del^{m}\!M^{*\bb} - P_{i\bj}\,\del_{m}A^i\del^m\!A^{*\bj}
- K_{a\bi} \del_m M^{a}\del^m\!A^{*\bi} - K_{\ba i}\del_m M^{*\ba} \del^m\!A^i\\
&  - G_{ab}\big(\tfrac{1}{4}\del_m B^a \del^m\!B^b + \tfrac{1}{6}C^a_{npq}C^{bnpq}\big) + \tfrac{i}{6} \big(G_{ai}\del_m A^i - G_{a\bj}\del_m A^{*\bj}\big) \veps^{mnpq}C^a_{npq}\\
& + \hat g_{ab}H^{a}H^{b} + \Im\l(\tilde Q_a R^{-1ab} K_{b \bc} - 2\tilde Z^*_\bj P^{-1\bj i} K_{i\bc} - 2  W_c\r) H^c - \mathcal V \ \bigg]\ , \eal
\end{equation}
where the 3-form field strengths come with the metric
\begin{equation}
 \hat g_{ab} = \Re \big(K_{ab} - K_{a\bj}P^{-1\bj i}K_{i\bb}\big) + (\Im K_{a\bar c})R^{-1cd}(\Im K_{d\bb})
\end{equation}
and the scalar potential is given by
\begin{equation}
 \mathcal{V} =  G_{ab} M^a M^{b*} + \tilde Z^*_{\bj}P^{-1\bj i}\tilde Z_i + \tfrac{1}{4}\tilde Q_{a} R^{-1ab}\tilde Q_b\ .
\end{equation}

\section{Conclusion}\label{sec:five}

In this paper we determined the couplings and dualities of
3-forms in $N=1, D=4$ globally supersymmetric theories. 
We gave the actions for massless and massive 3-form multiplets including
supersymmetric boundary terms.
We first discussed renormalizable interactions where we also allowed for the presence of a superpotential.
When dualizing these actions we found that a cosmological constant arises in the massless case
while the dual action of massive  3-forms contains an additional scalar field.
In analogy to the known duality between the chiral and complex linear multiplet, there appears a new multiplet that, in the massless case, differs from the complex linear multiplet only by a constant. 

In the non-renormalizable case we focused on the scalar geometry of
the non-linear sigma model for 3-form multiplets.
Elimination of the massless 3-forms from the action was demonstrated in order to find the on-shell action with the scalar potential.
We derived the Poincar\'e dual action and showed that in the massless case the
transition from action to dual action can be described by a field
redefinition so that the dual action comes with the same K\"ahler
geometry as the original one. 
In the massive case an additional real scalar appears and its 
sigma model metric is related to the  mass matrix of
the 3-forms.  In the dual action this scalar, together with the massive 3-form, is replaced by a complex scalar and the resulting geometry is the product of two K\"ahler manifolds.

Finally we coupled the massive 3-form multiplets to chiral multiplets,
studied the condition for supersymmetric backgrounds 
and determined a typical mass spectrum.


\vskip 1cm

\section*{Acknowledgments}
We would like to thank Vincente Cort\'es, Jim Gates and 
Gabriele Tartaglino-Mazzucchelli
 for useful  discussions and correspondence.
This work was supported by the German Science Foundation (DFG) under the
Collaborative Research Center (SFB) 676 ``Particles, Strings and the Early Universe''.
The research of K.~G.\ is supported by an ERC Starting Grant 277570-DIGT.

\vskip 2cm

\appendix

\noindent{\huge\bf Appendix}

\vskip 1cm

\section{Conventions} \label{conventions}

In this paper we use the conventions of \cite{WB}.
The Minkowski metric is $\eta = \tn{diag}(-1,1,1,1)$ and we fix the totally antisymmetric tensor $\veps^{mnpq}$ in four space-time dimensions by
\begin{equation}\label{epsilon}
 \veps^{0123} = 1\ , \qquad \veps_{0123} = \det g = -1 \quad \tn{for}\quad g = \eta\ .
\end{equation}
Spinor indices are raised and lowered with the antisymmetric tensor $\varepsilon^{\alpha\beta}$ as follows
\begin{equation}
\begin{aligned} \psi_{\alpha} = \veps_{\alpha\beta}\psi^{\beta}\ , \qquad \bar{\psi}_{\dal} = \veps_{\dal\dot{\beta}}\bar{\psi}^{\dot{\beta}}\ , \\
	\veps^{12}=\varepsilon_{21}=1\ , \quad \veps^{21}=\veps_{12}=-1\ , \quad \veps_{\alpha\beta}\veps^{\beta\gamma}&=\delta_{\alpha}^{\gamma}\ . \eal
\end{equation}
Two spinors can form Lorentz invariant products by contraction of their indices:
\begin{equation}
 \psi\chi = \psi^{\alpha}\chi_{\alpha} = \chi\psi\ , \qquad \bar{\psi}\bar{\chi} = \bar{\psi}_{\dal}\bar{\chi}^{\dal} = \bar{\chi}\bar{\psi}\ .
\end{equation}
The Pauli matrices $\sigma^m$ and $\bar\sigma^m$ are defined by
\begin{equation} \sigma^m_{\alpha\dal} := (-\mathbf{1}, \sigma^1, \sigma^2, \sigma^3)_{\alpha\dal}\ , \qquad  \bar{\sigma}^{m\dal\alpha} := \veps^{\dal\dot{\beta}}\veps^{\alpha\beta}\sigma^{m}_{\beta\dot{\beta}}\ ,
\end{equation}
and the conventions for the superspace integration are
\begin{equation}
\ivoll := \ifull\ , \qquad
    d^2\!\theta := \tfrac{1}{4} \veps^{\alpha\beta} d\theta_{\alpha}d\theta_{\beta}\ , \qquad
    d^2\!\bar\theta := -\tfrac{1}{4}\veps^{\dal\dot{\beta}}d\bth_{\dal}d\bth_{\dot{\beta}}\ .
\end{equation}
The generators of supersymmetry $Q_\alpha,\ \bar Q_{\dal}$ were chosen to be represented by
\begin{equation}
Q_{\alpha} = \der{}{\theta^\alpha} - i\sigma^{m}_{\alpha\dot{\alpha}}\bar{\theta}^{\dot{\alpha}}\del_m\ , \qquad
\bar{Q}_{\dot{\alpha}} = -\der{}{\btheta^{\dot\alpha}} + i\theta^{\alpha}\sigma^{m}_{\alpha\dot{\alpha}}\del_m\ ,
\end{equation}
while the covariant superspace derivatives $D_\alpha$, $\bar D_{\dot\alpha}$ are 
defined as
\begin{equation}
  D_\alpha = \der{}{\theta^\alpha} + i \sigma^m_{\alpha\dot\alpha}\btheta^{\dot\alpha}  \del_m\ , \qquad
 \bar D_{\dot\alpha} =  -\der{}{\btheta^{\dot\alpha}} - i\theta^{\alpha} \sigma^m_{\alpha\dot\alpha}  \del_m\ . 
\end{equation}

\section{N=1 supersymmetry multiplets} \label{multiplets}

For completeness  we recall various $N=1$ supermultiplets
in this appendix.

\subsection{Chiral multiplet} \label{chiral_multiplet}
The chiral superfield is defined by
\begin{equation}
  \bar{D}_{\dal}\Phi=0\ ,
\end{equation}
and can always be expressed via an unconstrained complex superfield $F$ as
$\Phi=\bar{D}^{2}F$. In terms of component fields it  has the generic form
\begin{equation}\begin{aligned} \label{chirmul}
	\Phi = \ &A + i\theta\sigma^{m}\bar{\theta}\del_{m}A +\tfrac{1}{4}\thvoll\Box A
	 +\sqrt{2}\theta\psi -\tfrac{i}{\sqrt{2}}\theta\theta\del_{m}\psi \sigma^{m}\bar{\theta}
	+\theta\theta F\ ,
\end{aligned}\end{equation}
containing a complex scalar $A$, a Weyl fermion $\psi$ and an auxiliary field $F$. Its renormalizable kinetic action is given by
\begin{equation}
\bal S = \ivoll \Phi \bar\Phi = \ix \big(-\del_mA\, \del^m\!A^{*} - i\psi\sigma^{m}\del_{m}\bar{\psi}+FF^{*}\big)\ . \eal
\end{equation}

\subsection{Vector multiplet}

The vector multiplet is represented by a real superfield $V = \bar V$. Its $\theta$-expansion can be written as
\begin{equation}\begin{aligned} \label{vectormulti}
	V =\ &B + i\theta\chi - i\bar{\theta}\bar{\chi} + \thz M^* + \thq M
+ 2 \theta\sigma^{m}\bar{\theta}v_{m}\\
&+\thz\bar{\theta} \big(\sqrt{2}\bar{\lambda} + \tfrac{1}{2}\bar{\sigma}^{m}\del_{m}\chi \big) + \thq\theta \big(\sqrt{2}\lambda - \tfrac{1}{2}\sigma^{m}\del_{m}\bar{\chi} \big)
	+\thvoll \big(D - \tfrac{1}{4}\Box B \big)\ ,
\end{aligned}\end{equation}
with real scalars $B$ and $D$, a complex scalar $M$, a real vector $v_{m}$ and Weyl spinors $\chi,\ \lambda$.
The vector multiplet is used for the description of supersymmetric gauge theories with $v_m$ being the gauge boson. A gauge transformation is implemented as
\begin{equation}\begin{aligned} \label{vector_gauge}
	V \to \ V+\Phi+\bar\Phi\ , \quad v_{m}  \to \ v_{m} + \tfrac{i}{2}\del_{m}(A-A^{*})\ ,
\end{aligned}\end{equation}
with a chiral superfield $\Phi$. The (Abelian) field strength multiplet of $V$,
 invariant under \eqref{vector_gauge}, is defined by
\begin{equation} \label{vector_fieldstrength}
	W_{\alpha} = -\tfrac{1}{4}\bar{D}^{2} D_{\alpha}V\ ,
\end{equation} 
and it contains the field strength $v_{mn} =
\del_{m}v_{n}-\del_{n}v_{m}$. With the help of
\eqref{vector_gauge} one can go to the Wess-Zumino gauge where  the
components $B$, $\chi$ and $M$ in \eqref{vectormulti} are set to zero.

\subsection{Linear multiplet} \label{ch_linearmultiplet}

A real multiplet $L= \bar L$ that satisfies the additional constraint \cite{GGRS}
\begin{equation}\label{Lcon}
 D^2 L = 0\ , \qquad \bar D^2 L = 0
\end{equation}
is called linear multiplet. Its component form is given by
\cite{massivetensor}
\begin{equation}
 \label{linearmulti}
L= E + i\theta\eta - i\bar{\theta}\bar{\eta} + \tfrac{1}{3}\thsigmath\veps_{mnpq}\del^{[n}B^{pq]}+\tfrac{1}{2}\thz\bar{\theta}\bar{\sigma}^{m}\del_{m}\eta-\tfrac{1}{2}\thq\theta\sigma^{m}\del_{m}\bar{\eta}-\tfrac{1}{4}\theta\theta\bar{\theta}\bar{\theta}\Box E\ ,
\end{equation}
containing a real scalar $E$, the field strength of a 2-form $B^{pq}$ and a Weyl spinor $\eta$. 
The action reads
\begin{equation}
	S = -\ivoll L^{2} 
	= \ix \big(-\tfrac{1}{2}\del_{m}E\del^{m}E -i\eta\sigma^{m}\del_{m}\bar{\eta}
	-\tfrac{1}{3}\del_{[n}B_{pq]}\del^{[n}B^{pq]}\big).
\end{equation}
The linear multiplet carries, like the chiral multiplet, $(4+4)$ 
(4 bosonic and 4 fermionic) degrees of freedom off-shell. On-shell it
has $(2+2)$ degrees of freedom and contains no auxiliary fields. There
is a duality between the chiral and linear multiplet that corresponds
to the on-shell equivalence of a scalar field and a 2-form. 

\subsection{Complex linear multiplet} \label{complex_linear}

The complex linear multiplet $\Sigma$ is defined by a condition similar to
\eqref{Lcon}
but with $\Sigma$  being complex \cite{nonmsca,comments}:
\begin{equation}
 D^{2}\bar\Sigma = 0\ , \qquad \bar{D}^{2}\Sigma = 0\ .
\end{equation}
This implies 
the component expansion
\begin{equation}\begin{aligned} \label{clm}
	\Sigma = \ &f + \theta\psi + \sqrt{2}\bar{\theta}\bar{\varphi} + \thz h + \thsigmath w_{m}
	+\thz\bar{\theta}\bar{\vartheta}-\tfrac{i}{\sqrt{2}}\thq\theta\sigma^{m}\del_{m}\bar{\varphi} \\
	&+\thvoll \big(-\tfrac{i}{2}\del_{m}w^{m}-\tfrac{1}{4}\Box
        f\big)\ , \\
\end{aligned}\end{equation}
where $f,h$ are complex scalars, $w_m$ is a complex vector and 
$\psi,\varphi,\vartheta$ are Weyl spinors. Altogether these are 
$(12+12)$ off-shell degrees of freedom.
The action for $\Sigma$ reads
\begin{equation}\begin{aligned} \label{S_clm}
	S = -\ivoll \Sigma\bar\Sigma
= &\ix \big(\tfrac{i}{2} f^{*}\del_{m}w^{m} -\tfrac{i}{2} f\del_{m}w^{m*} +\tfrac{1}{2}f\Box f^{*}
	+\tfrac{1}{2}\psi\vartheta+\tfrac{1}{2}\bar{\psi}\bar{\vartheta} \\
	 & \qquad\qquad -i\varphi\sigma^{m}\del_{m}\bar{\varphi} -
         hh^{*} + \tfrac{1}{2}w_{m}^{*}w^{m}\big)\ .
\end{aligned}\end{equation}
After elimination of the auxiliary fields $w_m,\ h,\ \vartheta$ and
$\psi$ 
one obtains the on-shell action 
\begin{equation}
 S = \ix \big(-\del_m f\, \del^m\!f^* - i\varphi\sigma^m \del_m \bar\varphi \big).
\end{equation}
Like the action of the chiral multiplet, it describes a complex scalar and a Weyl spinor. Therefore the chiral multiplet can alternatively  be dualized to a complex linear multiplet~\cite{nonmsca}.

\section{The massless 3-form action: boundary terms and duality} \label{app_3form}

In this appendix we discuss the action of a massless 3-form, its dualization and its connection to the cosmological constant. We deal with the issue of boundary terms and show that the appropriate variational constraint says
that the variation of the scalar field strength has to vanish at the boundary of the integration volume.

The canonical action of a massless 3-form is\footnote{The usual normalization includes another factor of 1/2 which we omit for convenience.} 
\begin{equation} \label{S_3}
 S_3 = - \tfrac{1}{24} \ix H_{mnpq}H^{mnpq}  = \ix H^2\ ,
\end{equation}
where  $H_{mnpq} = 4\del_{[m}C_{npq]} =  -\varepsilon_{mnpq}H$.
The equation of motion for the 3-form
\begin{equation}\label{HHrel}
   \veps^{mnpq} \del_m H = 0
\end{equation}
forces the field strength to be a constant, $H = c$ with $c\in \mathbb R$, or
\begin{equation}\label{sol_H}
 H^{mnpq} = - c \, \veps^{mnpq}\ .
\end{equation}
 For this reason the massless 3-form has been studied in the context of the
problem of the cosmological constant 
\cite{Hawking, Duff, Wu, Duncan, Bousso:2000xa}. 

However, the action \eqref{S_3} is not the full story since its variation
includes a boundary term of the form
\begin{equation}
 \delta S_3 =  \tfrac{1}{3} \ix \del_m \big(H \veps^{mnpq} \delta C_{npq}\big) - \tfrac{1}{3} \ix  (\del_m H) \veps^{mnpq} \delta C_{npq}\ .
\end{equation}
Thus, for the action \eqref{S_3} one has to impose 
\begin{equation} \label{VC_C}
 \delta C_{npq}\big\vert_{\del \mathcal M} = 0\ ,
\end{equation}
in order to make the boundary term vanish ($\del \mathcal{M}$ denotes the boundary of the integration volume $\mathcal{M}$).\footnote{Alternatively one could demand $\delta C^{npq}(x) \to 0$ for $x \to\infty$ sufficiently fast when one integrates over the whole Minkowski space.} One might already doubt that \eqref{VC_C}
is a valid boundary condition as it is not gauge invariant.
Moreover, it has been pointed out in ref.~\cite{Duff} that substituting the solution \eqref{sol_H} back into \eqref{S_3} yields the wrong sign for the correction of the bare cosmological constant $\Lambda_0$. The correct value of the effective cosmological constant can be found by coupling the 3-form to gravity
via the action
\begin{equation} \label{S_3_EH}
 S_{3,\tn{EH}} = \tfrac{1}{16\pi G} \ix \sqrt{-g}(R-2\Lambda_0) + \ix
        \sqrt{-g} H^2\ , 
\end{equation}
where 
\begin{equation}
H = \tfrac{1}{24\sqrt{-g}}\,\veps^{mnpq}H_{mnpq}\ , \quad H_{mnpq} = -\tfrac{1}{\sqrt{-g}}\, \veps_{mnpq} H\ .
\end{equation}
(Here we used \eqref{epsilon} which implies for example 
$\veps^{mnpq}\veps_{mnpq} = 24g$ and the definition of $H$ was chosen such that it is a Lorentz scalar.)

The equation of motion of the 3-form again reads
$\veps^{mnpq}\del_m H = 0$
which is solved by
\begin{equation} \label{sol_H_grav}
 H = c\ ,\quad \tn{or} \quad \sqrt{-g} = \tfrac{1}{24c}\veps^{mnpq}H_{mnpq}\ .
\end{equation}
Inserting this solution into the stress energy tensor 
\begin{equation}
  T^{mn} =  -g^{mn} H^2
\end{equation}
which appears in the Einstein equations $R^{mn} - \frac{1}{2}g^{mn}R = -\Lambda_0 g^{mn} + 8\pi GT^{mn}$,
one computes an effective cosmological constant $\Lambda = \Lambda_0 + 8\pi G c^2$. 
On the other hand, substituting \eqref{sol_H_grav} into the action \eqref{S_3_EH} yields 
 $\Lambda = \Lambda_0 - 8\pi G c^2$.
This discrepancy is clearly 
a result of the incompatibility of the variational constraint
\eqref{VC_C} with the solution \eqref{sol_H_grav}. More precisely,  in
order to implement \eqref{VC_C} in the on-shell action,
\eqref{sol_H_grav} constraints
the variations of the metric by
\begin{equation}
 \int_{\mathcal M} d^4\!x\; \delta \sqrt{-g} = \tfrac{1}{24c} \int_{\mathcal M} d^4\!x\; \del_m(\veps^{mnpq} \delta C_{npq})  = 0\ ,
\end{equation}
which is not a reasonable constraint. In fact, there is no way at all to implement the constraint $\delta C_{npq}\vert_{\del \mathcal M} = 0$ in the on-shell action since for given $\delta g_{mn}$ the solution \eqref{sol_H_grav} fixes $\delta C_{npq}$ only up to a gauge transformation.  Thus it is not possible to derive a consistent on-shell action from the action \eqref{S_3_EH}. To cure this, one
imposes a different variational constraint on the 3-form
by  demanding \cite{Duncan} 
\begin{equation} \label{VC_H}
 \delta H \big\rvert_{\del \mathcal M} = 0\ .
\end{equation}
This condition is automatically fulfilled by \eqref{sol_H_grav}. In order to apply \eqref{VC_H}
one modifies the action \eqref{S_3} by adding the following boundary term \cite{Duncan}  
\begin{equation} \label{S_3_new}
 S^{\prime}_3 =   \ix  H^2 - \tfrac{1}{3} \ix  \del_m \big(H \veps^{mnpq}  C_{npq}\big)\ ,
\end{equation}
which does not alter the equations of motion.\footnote{Interestingly, this boundary term formally breaks the gauge invariance of the 3-form action.}
Indeed, the variation of \eqref{S_3_new} is given by
\begin{equation}
 \delta S_3^{\prime} =  -\tfrac{1}{3} \ix  (\del_m H) \veps^{mnpq} \delta C_{npq} - \tfrac{1}{3} \ix \del_m \big(\delta H \veps^{mnpq} C_{npq}\big)\ .
\end{equation}
Substituting the solution \eqref{sol_H} into $S_3 ^{\prime}$, we find that the boundary term has the opposite sign than the kinetic term and 
is twice as big resulting in
\begin{equation}
 \langle S_3^{\prime} \rangle = c^2 - 2c^2 = -c^2\ ,
\end{equation}
which is indeed the correct positive contribution to the cosmological constant.\footnote{Ref.~\cite{Duncan} also argued in the context of the Baum-Hawking-Coleman mechanism that adding the boundary term is necessary in order to get the right behavior of the quantity $\exp(-\langle S_{3,\tn{EH}} \rangle)$ which is maximized for $\Lambda \rightarrow 0_+$ and thus could provide a statistical explanation for the smallness of the cosmological constant.}

Finally, let us discuss the dual action of  \eqref{S_3_new}
including the boundary term. One couples the scalar field strength $H$ to another real scalar $\phi$ via the first-order action 
\begin{equation}\label{Sfirst}
\bal S_\tn{first} &= \ix \big(\! -\phi^2 + 2\phi H \big) - \tfrac{1}{3} \ix \del_m\big( \phi\,\veps^{mnpq} C_{npq}) \\
    &= \ix \big(\! -\phi^2 - \tfrac{1}{3} (\del_m \phi)\veps^{mnpq} C_{npq} \big)\ . \eal
\end{equation}
The equation of motion for $\phi$  is $\phi = H$ which, when
inserted into \eqref{Sfirst}, reproduces \eqref{S_3_new} correctly, including the boundary term. On the other hand, the equation of motion for  $C_{npq}$
constrains $\phi$ to be a constant, $\phi = c$ with $c \in \mathbb R$. Like in the original action \eqref{S_3_new}, the boundary term ensures
these equations without imposing that $\delta C_{npq}$ should vanish on the boundary. Here in the first order action the necessity of adding a boundary term becomes even more obvious, as it allows for the elimination of the 3-form from the action, leading to the dual action\footnote{Variations of the dual action with respect to the constant field $c$ are not allowed since we impose the constraint $\delta c\vert_{\del \mathcal M} = 0$, i.e., $\delta c = 0$.}
\begin{equation}
  S_\tn{dual} = \ix \l(-c^2\r)\ .
\end{equation}

\section{Elimination of auxiliary fields} \label{elim_aux}

Most of the known supermultiplets feature auxiliary fields that do not
correspond to on-shell degrees of freedom. These can be eliminated
from the action by using their purely algebraic equations of motion. 
The elimination of multiple auxiliary fields can become computationally
involved, especially for complex fields. However, when they occur 
quadratically in the action, a generalization of the well
known technique of 
``completing the square'' simplifies this task a lot.
Since we use this technique numerous times in the main text, we outline the general procedure in this appendix.

Suppose we have $N$ real auxiliary fields $D^a$ that occur in the Lagrangian as
\begin{equation} \label{L_real}
 \Lag = D^a M_{ab} D^b + J_a D^a + C\ ,
\end{equation}
where $M_{ab},\ J_a$ and $C$ are arbitrary functions of all other fields contained in the action. In order for the Lagrangian to be real, $J_a$ and $C$ have to be real and $M$ has to be a hermitian matrix, even though only its symmetric, 
i.e.\ real part contributes in \eqref{L_real}, which we take to be
invertible here.\footnote{If $\Re M$ was not invertible, each of its
zero eigenvalues would account for a constraint on the fields that
couple to the $D^a$ which reads $v^b J^b = 0$, where $v$ is a corresponding zero eigenvector.} To eliminate the $D^a$, we could simply insert their equations of motion
\begin{equation}
 2\Re (M_{ab}) D^b + J_a = 0 \qquad \=>\qquad D^b = -\tfrac{1}{2}(\Re M)^{-1ba} J_a
\end{equation}
into the Lagrangian \eqref{L_real}.  However the same can be achieved in a more elegant way by shifting the $D^a$,
\begin{equation}
 \tilde D^a := D^a + \tfrac{1}{2}(\Re M)^{-1ac} J_c\ ,
\end{equation}
to have the Lagrangian \eqref{L_real} assume the form
\begin{equation} \label{L_real_square}
 \Lag = \tilde D^a (\Re M)_{ab} \tilde D^b  - \tfrac{1}{4} J_a (\Re M)^{-1ab} J_b   + C\ .
\end{equation}
Now we immediately see that the first term in \eqref{L_real_square}
(the ``square'') vanishes by the equations of motion for the $D^a$ (or
$\tilde D^a$ respectively) and we can easily read off the final
Lagrangian. 

Now suppose that there are complex auxiliary fields $F^a$ that occur in the Lagrangian as
\begin{equation} \label{L_compl_1}
 \Lag = F^a K_{a\bb}F^{*\bb} + J_a F^a + J_\bb^*F^{*\bb} + C\ ,
\end{equation}
where $K$ is an invertible hermitian matrix, $J_a$ is a complex and $C$ a real function of the other fields. Here the square is completed by shifting
\begin{equation}
 \tilde F^a := F^a + J_{\bar c}^*K^{-1\bar c a}\ , 
\end{equation}
so that the Lagrangian becomes
\begin{equation}
 \Lag = \tilde F^{a} K_{a\bb} \tilde F^{*\bb} - J_{\ba}^*K^{-1\ba b}J_{b} + C\ .
\end{equation}
Again the ``square'' vanishes by the equations of motion for the $F^a$.

Note that \eqref{L_compl_1} does not give the most general form of quadratic
terms for $N$ complex auxiliary fields. One could also have terms
proportional to $FF$ and $F^*F^*$ multiplied by a symmetric matrix $M$
and its complex conjugate, i.e.\
\begin{equation} \label{L_compl_2}
 \Lag = F^a M_{ab} F^b + F^{*\ba}M^*_{\ba\bb}F^{*\bb} + 2F^{a}K_{a\bb}F^{*\bb} + J_a F^a + J_{\bb}^*F^{*\bb} + C\ .
\end{equation}
Now the task of completing the square is more complicated than for \eqref{L_compl_1}. However, by making the ansatz
\begin{equation} \label{L_compl_square_2}
\Lag = \big((F + T)^T \ (F + T)^\dagger\big) \begin{pmatrix} M & K \\ K^* & M^* \end{pmatrix} \begin{pmatrix} F + T \\ F^* + T^* \end{pmatrix} - \big(T^T \ T^\dagger\big) \begin{pmatrix} M & K \\ K^* & M^* \end{pmatrix} \begin{pmatrix} T \\ T^* \end{pmatrix} + C\ ,
\end{equation}
one finds that the $T^a$ have to satisfy
\begin{equation}
 2 \begin{pmatrix} M & K \\ K^* & M^* \end{pmatrix} \begin{pmatrix} T \\ T^* \end{pmatrix} = \begin{pmatrix} J \\ J^* \end{pmatrix}\ .
\end{equation}
Provided that $K$ and the matrix
\begin{equation}
 H := \begin{pmatrix}  M & K \\ K^* & M^* \end{pmatrix}
\end{equation}
are invertible, the inverse is of the form
\begin{equation} \label{H_inv}
 H^{-1} = \begin{pmatrix}  N & G \\ G^* & N^* \end{pmatrix}\ , \qquad \tn{where} \qquad 
\bal G &= \l(K^* - M^*K^{-1}M\r)^{-1}, \\  N &= -\l(K^{-1}MG\r)^*\ . \eal
\end{equation}
Then the on-shell Lagrangian becomes (note that $G$ is hermitian)
\begin{equation}
\bal
 \Lag_\tn{on-shell} &= - \tfrac{1}{4} \big(J^T \ J^\dagger \big) H^{-1} \begin{pmatrix} J \\ J^* \end{pmatrix} + C\\
    &= -\tfrac{1}{4} \big(J_a N^{ab} J_b + J^*_{\ba}N^{*\ba\bb}J^*_{\bb}\big) - \tfrac{1}{2} J_a G^{a\bb} J^*_{\bb} + C\ .
\eal
\end{equation}

\section{Dual sigma model actions with a shift symmetry}\label{app:real}

Not every sigma model action with 3-form multiplets can be dualized in the way described in sections \ref{sigma_dual_massl} and \ref{sigma_dual_massive}.
As an important application for string theory let us consider
the specific class of K\"ahler potentials with a shift symmetry
where $K$  only depend on the real parts of the $S^a$, i.e.
\begin{equation}
 K(S,\bar S) = K(S + \bar S) \quad \=> \quad \der{K}{S^a} = \der{K}{\bar S^a}\ . 
\end{equation}
Then the arguments $F_a$ of the Legendre transform $\hK$ of $K$ also have to be real and a first order action is, in the massless case, given by
\begin{equation} \label{S_first_real}
 S_{\tn{first}} = \ivoll \big( - \hK(F) + F_a(S^a + \bar S^a) \big) +
 \mathcal{B}_\tn{first}\ ,
\end{equation}
with boundary terms as before (cf. \eqref{B_first})
\begin{equation}
  \mathcal{B}_\tn{first} = \tfrac{1}{4} \ivoll \big[\bar D_\dalpha \l(F_a \bar D^\dalpha U^a - \bar D^\dalpha F_a  U^a\r) + \hc \big]\ .
\end{equation}
Since the $F_a$ are real, their component expansion can be written as
\begin{equation}
 F_{a} = f_{a} + \thz n_{a} + \thq n^{*}_{a} + \thsigmath w_{a m} + \thvoll \big(d_{a}-\tfrac{1}{4}\Box f_{a}\big)\ ,
\end{equation}
where $f_a, d_a$ and $w_{am}$ are real. The superfield equations of motion for the action \eqref{S_first_real} are
\begin{equation} \label{eom_S_first}
 S^a + \bar S^a = \der{\hK}{F_a}\ , \qquad (D^2 + \bar D^2) F_a = 0\ .
\end{equation}
The second equation (which is used to eliminate the 3-form multiplets from the action and find a dual action) imposes the constraints
\begin{equation}
 d_a = 0, \qquad \del_m w_a^m = 0, \qquad n_a = i\cc_a,\quad \cc_a \in \mathbb{R}\ 
\end{equation}
on the components of $F_a$. The second condition is solved by $w_{am} = \veps_{mnpq}\del^{[n}B_a^{pq]}$ with a 2-form $B^{pq}$. Therefore we find as a dual action
\begin{equation} \label{S_dual_real}
 \bal S_{\tn{dual}} &= - \ivoll \hK(F) 
    &= \ix \hK^{ab} \big( -\tfrac{1}{4}\del^{m}\!f_{a}\,\del_{m}f_{b} -\tfrac{3}{2}\del^{[n}B_{a}^{pq]}\del_{n}B_{b pq} -\cc_{a}\cc_{b} \big)\ . \eal
\end{equation}
The $2N_3$ bosonic on-shell degrees of freedom contained in the $M^a$
are distributed in the dual action among the real scalars $f_a$
and the 2-forms $B_a^{pq}$, each with $N_3$
degrees of freedom.

As an example consider the K\"ahler potential
\begin{equation}
 K(S, \bar S)\ =\ -\log(S + \bar S)\ .
\end{equation}
The massless action including the boundary term for the 3-form is then given by
\begin{equation}
 S = \ix \frac{1}{(2\Re M)^2}\Big(-\del_m M\, \del^m\! M^* + H^2\Big) -\tfrac{1}{3} \ix \del_m \Big(\frac{1}{(2\Re M)^2} H \veps^{mnpq} C_{npq} \Big)\ .
\end{equation}
To find  $\hK(F)$, use the Legendre relation $F = \del K/\del S = -(S + \bar S)^{-1}$ and
\begin{equation}
 \hK(F)\ =\ -K(S, \bar S) + F(S + \bar S)\ =\ \log(-F) - 1\ .
\end{equation}
Thus the dual acion is given by
\begin{equation}
 \bal S_{\tn{dual}} = - \ivoll \hK(F) 
    = \ix \frac{1}{f^2} \big(-\tfrac{1}{4}\del^{m}\!f\,\del_{m}f - \tfrac{3}{2}\del^{[n}B^{pq]}\del_{n}B_{pq} - \cc^2 \big)\ . \eal
\end{equation}

\section{Legendre transformation} \label{App_Legendre}

In this appendix we assemble a few facts about the Legendre transformation.
For a smooth, strictly convex function $K: \ \mathbb{R}^n \to \mathbb{R}$ the gradient function
\begin{equation}\label{s_x_relation}
 s:\ \mathbb{R}^n \to \mathbb{R}^n,\quad  s_i(x) := \der{K}{x^i}(x)
\end{equation}
is invertible \cite{legendretrans} and the inverse is denoted by $x(s)$. 
Then the Legendre transform of $K$ is defined by 
\begin{equation} \label{legendredef}
	\hK(s) := s_i x^i(s) - K(x(s))\ .
\end{equation}
In particular, $\hK$  satisfies
\begin{equation}
  \hK(\der{K}{x})\ =\ \der{K}{x^i} x^i - K(x) \quad \tn{for all}\ x \in  \mathbb{R}^n\ .
\end{equation}
For the derivative of $\hK$ one finds
\begin{equation} \label{der_hK}
 \der{\hK}{s_i}(s)\ = \ s_j\,\der{x^j}{s_i} + x^i - \der{K}{x^j}(x(s))\der{x^j}{s_i}\ =\ x^i(s)\ .
\end{equation}
If we denote the variable of the double Legendre transform $\hat{\hK}$ as $\tilde x$, the function $s(\tilde x)$ is defined by
\begin{equation}
 \der{\hK}{s_i}(s(\tilde x)) = \tilde x^i\ .
\end{equation}
Thus equation \eqref{der_hK} shows that $x(s(\tilde x)) = \tilde x$ 
(i.e., $\tilde x$ is really the original variable $x$) which implies that the Legendre transformation is its own inverse:
\begin{equation}
\bal \hat{\hK}(\tilde x)\ &= \ \tilde{x}^i s_i(\tilde x) - \hK(s(\tilde x))\\
    &= \ \tilde{x}^i s_i(\tilde x) - \big(s_i(\tilde x) x^i(s(\tilde x)) - K(x(s(\tilde x)))\big)\ = \ K(\tilde x)\ . \eal
\end{equation}

From the relations \eqref{s_x_relation} and \eqref{der_hK} it follows 
that 
\begin{equation}
\dder{K}{x^i}{x^j}\,\dder{\hK}{s_j}{s_k} = \der{s_i}{x^j}\,\der{x^j}{s_k} = \delta_i^k\ , \qquad {\rm or}\qquad
\Hess K = \big( \Hess \hK \big)^{-1} \ ,
\end{equation}
where the derivatives of $\hK$ have to be evaluated at $s(x) = \del K/\del x$ when those of $K$ are evaluated at $x$.

In the case of a K\"ahler potential $K(z, \bar z)$ with K\"ahler metric
\begin{equation}
 K_{i\bj} = \dder{K}{z^i}{\bar z^j}
\end{equation}
one has
\begin{equation} \label{Hesse}
 \Hess K = \begin{pmatrix}
  K_{ij} & K_{i\bj} \\
  K_{\bi j} & K_{\bi\bj}
 \end{pmatrix}, \qquad  \Hess \hK = \begin{pmatrix} \hK^{ij} & \hK^{i\bj} \\ \hK^{\bi j} & \hK^{\bi\bj}
 \end{pmatrix}\ .
\end{equation}
The inverse of a block matrix of the form of $\Hess \hK$ is given in equation \eqref{H_inv}. Thus we obtain the formula
\begin{equation} \label{kaehler_metric}
 K_{i\bj} = \l(\hK^{\bj i} -  \hK^{\bj \bar{k}}\hK^{-1}_{\bar{k}l}\hK^{li}\r)^{-1}\ .
\end{equation}

\bibliographystyle{plain}

\end{document}